\documentclass[11pt]{article}
\usepackage{amsmath,epsf,amssymb,latexsym,enumerate,cite,shadow,array,color,bbold,slashed}
\usepackage{ifpdf}
\ifx\pdfoutput\undefined
   \pdffalse
   \usepackage{cite}
 \else
   \pdfoutput=1
   \pdftrue
  \usepackage[pdftex]{hyperref}
\fi
\newsavebox{\uuunit}
\sbox{\uuunit}
    {\setlength{\unitlength}{0.825em}
     \begin{picture}(0.6,0.7)
        \thinlines
        \put(0,0){\line(1,0){0.5}}
        \put(0.15,0){\line(0,1){0.7}}
        \put(0.35,0){\line(0,1){0.8}}
       \multiput(0.3,0.8)(-0.04,-0.02){12}{\rule{0.5pt}{0.5pt}}
     \end {picture}}
\newcommand {\unity}{\mathord{\!\usebox{\uuunit}}}

\csname @addtoreset\endcsname{equation}{section}

\DeclareFontFamily{U}{rsf}{}
\DeclareFontShape{U}{rsf}{m}{n}{
  <5> <6> rsfs5 <7> <8> <9> rsfs7 <10-> rsfs10}{}
\DeclareMathAlphabet\Scr{U}{rsf}{m}{n}

\def\pplogo{\vbox{\kern-\headheight\kern -29pt
\halign{##&##\hfil\cr&{
\ppnumber}\cr\rule{0pt}{2.5ex}&\ppdate\cr}
}}
\makeatletter
\def\ps@firstpage{\ps@empty \def\@oddhead{\hss\pplogo}%
  \let\@evenhead\@oddhead 
}
\def\maketitle{\par
 \begingroup
 \def\thefootnote{\fnsymbol{footnote}}
 \def\@makefnmark{\hbox{$^{\@thefnmark}$\hss}}
 \if@twocolumn
 \twocolumn[\@maketitle]
 \else \newpage
 \global\@topnum\z@ \@maketitle \fi\thispagestyle{firstpage}\@thanks
 \endgroup
 \setcounter{footnote}{0}
 \let\maketitle\relax
 \let\@maketitle\relax
 \gdef\@thanks{}\gdef\@author{}\gdef\@title{}\let\thanks\relax}
\makeatother

\makeatletter
\renewcommand\section{\@startsection {section}{1}{\z@}%
                                   {-3.5ex \@plus -1ex \@minus -.2ex}%
                                   {2.3ex \@plus.2ex}%
                                   {\normalfont\large\bfseries}}
\renewcommand\subsection{\@startsection{subsection}{2}{\z@}%
                                   {-3.25ex\@plus -1ex \@minus -.2ex}%
                                   {1.5ex \@plus .2ex}%
                                   {\normalfont\normalsize\bfseries}}

\newcommand{\ft}[2]{{\textstyle\frac{#1}{#2}}}

\def\Tr{\mathop{\rm Tr}\nolimits}
\def\rmi{{\rm i}}
\def\rmd{{\rm d}}

\newcommand{\cG}{{\cal G}}
\newcommand{\betaplus}{\beta _+}
\newcommand{\betaminus}{\beta _-}
\newcommand{\betaplusprime}{\beta }
\newcommand{\cN}{{\cal N}}

\begin{document}

\begin{titlepage}
\

\vspace{0.5cm}

\centerline{\Large \bf Dirac-Born-Infeld-Volkov-Akulov   and Deformation
of Supersymmetry}

 \vskip 15mm 
\begin{center}{\large \bf Eric Bergshoeff$^1$, Frederik Coomans$^2$, Renata Kallosh$^3$,\\
C.~S.~Shahbazi$^{4}$ and Antoine Van Proeyen$^2$} \\[1.5truecm]

{\small  $^1$ Centre for Theoretical Physics, University of Groningen,
Nijenborgh 4,\\
  9747 AG Groningen, The Netherlands \\[.2truecm]
  $^2$
Instituut voor Theoretische Fysica, Katholieke Universiteit  Leuven,\\
       Celestijnenlaan 200D B-3001 Leuven, Belgium\\[.2truecm]
  $^3$ Stanford Institute for Theoretical Physics and Department of Physics,\\
  Stanford University, Stanford, CA 94305\\[.2truecm]
 $^4$ Instituto de F{\'\i}sica Te{\'o}rica UAM/CSIC
\\
C/ Nicol{\'a}s Cabrera, 13--15,  C.U.~Cantoblanco, 28049 Madrid, Spain}

\vskip 1truecm

{\bf Abstract}
\end{center}

We deform the action and the supersymmetry transformations of  the
$d=10$ and $d=4$  Maxwell supermultiplets so that at each order of the
deformation the theory has 16 Maxwell multiplet deformed supersymmetries
as well as 16 Volkov-Akulov type non-linear supersymmetries. The result
agrees with the expansion in the string tension of the explicit action
of  the Dirac-Born-Infeld model and its supersymmetries, extracted from
D9 and D3 superbranes, respectively. The half-maximal Dirac-Born-Infeld
models with 8 Maxwell supermultiplet deformed  supersymmetries and 8
Volkov-Akulov type supersymmetries are described by a new class of $d=6$
vector branes related to chiral (2,0) supergravity, which we denote as
`Vp-branes'. We use a space-filling V5 superbrane for the $d=6$ model
and a V3 superbrane for the $d=4$ half-maximal Dirac-Born-Infeld (DBI)
models. In this way we present a completion to all orders of the
deformation of the Maxwell supermultiplets with maximal 16+16
supersymmetries in $d=10$ and 4, and half-maximal 8+8 supersymmetries in
 $d=6$ and 4.

 \vfill

\hrule width 3.cm \vspace{2mm}{\footnotesize \noindent e-mails:
E.A.Bergshoeff@rug.nl, \{Frederik.Coomans,
Antoine.VanProeyen\}@fys.kuleuven.be, \\[-2mm] \phantom{e-mails:}kallosh@stanford.edu,
carlos.shabazi@uam.es}

\end{titlepage}
\addtocounter{page}{1}
\tableofcontents
\newpage

\section{Introduction}
The purpose of this study is to look for new ways of constructing
supersymmetric invariants for theories with extended supersymmetries
where there are no known auxiliary fields. For example, the
supersymmetries of a $d=4$, $\cN=4$ Maxwell multiplet form an open
algebra, it is closed only when the classical equations of motion are
satisfied. This prevents  the direct use of the $\cN=4$ superconformal
tensor calculus to construct superconformal invariants with higher
derivatives \cite{Ferrara:2012ui}. It is different from the $\cN=2$
case, where some superconformal multiplets form a closed algebra, the
auxiliary fields are known, and  higher derivative superconformal
invariants can be constructed using superconformal calculus, as shown in
detail in \cite{deWit:2010za,Chemissany:2012pf}. When auxiliary fields
are eliminated using their equations of motion, one finds a deformed
local $\cN=2$ supersymmetry and the deformed $\cN=2$ supergravity action
(after gauge fixing of extra symmetries), which depend only on physical
fields \cite{Chemissany:2012pf}.

Here we will build the higher derivative supersymmetric gauge theory
model developing the proposal in \cite{Bergshoeff:1986jm} to deform the
quadratic action of the $d=10$ Maxwell multiplet where the deformation
parameter is the open string tension. We refer the reader to Born-Infeld
and  Dirac models  \cite{Born:1934gh,Dirac:1962iy},
its supersymmetric generalizations  and its  relation to string theory
discussed in
\cite{Deser:1980ck,Cecotti:1986gb,Metsaev:1987qp,Bagger:1996wp,Rocek:1997hi,Tseytlin:1999dj,Ketov:1998ku,Bellucci:2000ft,Ketov:2001dq,Kuzenko:2000tg,Kuzenko:2000uh,Bellucci:2001hd,Kerstan:2002au,Ivanov:2002ab,Berkovits:2002ag,Ivanov:2003uj}.
A superembedding approach as a generic covariant method for the
description of superbranes as models of partial spontaneous
supersymmetry breaking was developed in
\cite{Sorokin:1999jx,Pasti:2000zs}. In the context of extended $d=4$
supergravity and duality symmetry there is a significant interest to
Born-Infeld type constructions
\cite{Carrasco:2011jv,Chemissany:2011yv,Broedel:2012gf,Pasti:2012wv,Ivanov:2012bq,Kuzenko:2013gr,Aschieri:2013nda}.

Our work will consist of  bottom up and top down deformation of the
supersymmetric Maxwell action:
\begin{equation}\label{Maxwellaction}
S = \int \rmd^{10}x\,\big \{-\tfrac{1}{4} (F_{\mu\nu})^2 + \bar\lambda \slashed{\partial} \lambda \big\}\,.
\end{equation}
where $F_{\mu\nu}= \partial_\mu A_\nu - \partial_\nu A_\mu$. There is no
known off-shell formulation of this multiplet. The same applies to the
$\cN=4$ version of it in $d=4$ where our current goal is to learn some
new ways of building higher derivative actions involving this
supermultiplet \cite{Ferrara:2012ui}.

This theory describes $8+8$ on-shell degrees of freedom and consists of a vector $A_\mu$ and a 16-component Majorana-Weyl spinor $\lambda$.
The on-shell 16 linear supersymmetries  are given by
\begin{equation}
\delta_\epsilon  A_\mu = {\bar\epsilon}\Gamma_\mu\lambda \,,\qquad
\delta_\epsilon \lambda = \tfrac{1}{4}\Gamma^{\mu \nu}F_{\mu \nu}\epsilon\,.
\label{epsilon}
\end{equation}
The action (\ref{Maxwellaction}) has also a trivial  fermionic symmetry
under which the fermion shifts by a  constant parameter
\begin{equation}
\delta_\eta  A_\mu = 0\,,\qquad  \delta_\eta \lambda= -\tfrac{1}{2\alpha }\eta\,.
\label{eta}
\end{equation}

In the bottom up procedure we will  deform  the Maxwell action  and the
16 supersymmetries of the linear action order by order in the string
tension, so that at each order the deformed action has the symmetries
(\ref{epsilon}), deformed by ${\cal O}(\alpha)$ terms, as well as 16
hidden Volkov-Akulov type supersymmetries \cite{Volkov:1973ix} of the
form
\begin{eqnarray}
\delta_{\zeta} A_\mu= -\bar\zeta
 \Gamma_\mu
\lambda
+ {\cal O} (\alpha)\, , \qquad
\delta_{\zeta} \lambda = \alpha^{-1}\zeta +{\cal O} (\alpha)\,.
\label{AsusyShort}
\end{eqnarray}
We will compare this with the top down approach based on the D9
super-brane action,
\cite{Cederwall:1996pv,Cederwall:1996ri,Bergshoeff:1996tu,Aganagic:1996nn,Bergshoeff:1997kr,Kallosh:1997aw}.
It has been noticed in  \cite{Tseytlin:1999dj} that the supersymmetry of
the D9 super-brane action, upon gauge-fixing of a local
$\kappa$-symmetry, has a complicated form. This may have been one of the
reasons that in $d=4$ a better understanding of Dirac-Born-Infeld (DBI)
type models was achieved in the past by developing the superfield models
where, for example, the linear $\cN=2$ off shell supersymmetry including
auxiliary fields was manifest, whereas the non-linear spontaneously
broken one was deformed and often called `hidden'
\cite{Bagger:1996wp,Rocek:1997hi,Tseytlin:1999dj,Ketov:1998ku,Ketov:2001dq,Kuzenko:2000tg,Kuzenko:2000uh,Bellucci:2001hd,Sorokin:1999jx,Pasti:2000zs,Carrasco:2011jv,Chemissany:2011yv,Broedel:2012gf}.

Here we will present an explicit,  relatively simple and complete form
of the $d=10$ DBI model with 16 deformed supersymmetries of the Maxwell
multiplet and 16 Volkov-Akulov (VA) type non-linear supersymmetries. The
original VA model was proposed in $d=4$ in \cite{Volkov:1973ix}. The
$d=10$ analog was discussed in the context of the D9 branes in
\cite{Kallosh:1997aw}. The details of the derivation of this model from
the $\kappa$-symmetric D9-superbrane action are presented in  Appendix
\ref{app:A}. It will also be shown that  all these symmetries of the
DBI-VA model can be expanded in the string tension deformation parameter
and this expansion coincides with the bottom up model of the deformation
of the Maxwell multiplet. Thus, there is a {\it known completion of the
deformation process} and the action as well as all non-linear $16+16$
supersymmetries can be given in the compact form.\footnote{We will use
the terminology $16+16$ supersymmetry. The first term `16' refers to the
number of supersymmetries that have a linear part and imply equal number
of bosons and fermions and existence of the particle representations.
The second term `16' refers to non-linear supersymmetries of the VA
type, that do not determine the particle content. The same holds for
$8+8$ case.} Using the D3 super-brane we will also present a complete
and explicit $d=4$ DBI model with maximal number of non-linear
supersymmetries, half being VA-type. This new model with 16+16
supersymmetries in $d=4$ could be called $\cN=8$ DBI, by analogy with
the $\cN=4$ DBI model in \cite{Bellucci:2001hd} with 8+8
supersymmetries. No such DBI model was constructed before.

To study the half-maximal supersymmetric DBI models with 8  deformed
supersymmetries and 8 VA-type supersymmetries of the Maxwell multiplet,
we will introduce a new class of $d=6$ `vector branes'  whose
world-volume dynamics is described by a vector multiplet but whose
tension does not necessarily scale with the inverse string coupling
constant, like it is the case for Dirichlet branes. Such branes are
suggested by a recent analysis of the  different branes, and their
world-volume content, of the $d=6$ half-maximal theories
\cite{Bergshoeff:2012jb}. For our purposes it is sufficient to consider
the vector branes related to $d=6$ (2,0) chiral supergravity.

As inherited from the supersymmetric DBI model in $d=10$ and $d=6$, the
supersymmetric DBI models in $d=4$  feature a complete and explicit
deformation to all orders of all supersymmetries. This might help to
provide an all order  completion of the $\cN=4$ DBI model in
\cite{Bellucci:2001hd} with the same number of supersymmetries.

To compare our new complete models with the ones in $d=4$,
\cite{Cecotti:1986gb,Bagger:1996wp,Rocek:1997hi,Tseytlin:1999dj,Ketov:1998ku,Ketov:2001dq,Kuzenko:2000tg,Kuzenko:2000uh,Bellucci:2001hd},
will require a separate investigation. In these models, half of the
supersymmetry is manifest in superspace; it involves the   auxiliary
fields and it is not deformed order by order (only the hidden
supersymmetry is deformed). The action of these models with 8
supersymmetries manifest and 8  hidden supersymmetries is known only up
to order 10 in superfields, as shown in \cite{Bellucci:2001hd}. There
seems to be no known algorithm  which would generate the action at
higher orders of deformation. The actions which we will construct here,
with the 16+16 and 8+8  all deformed non-linear supersymmetries,  are
complete and therefore will be known at every level of deformation,
however none of the supersymmetries will be manifest.

This paper is organized as follows. In section \ref{ss:BIMaxd10D9} we
present the DBI action with 16+16 supersymmetries starting from the D9
superbrane. We do both a bottom-up and top-down calculation and compare
the two approaches. In section \ref{ss:16d4fromD3} we perform a similar
calculation to obtain the maximal ${\cal N}=4$ + ${\cal N}=4$ DBI action
as it follows from the D3 superbrane. The vector branes, that are
relevant for a brane interpretation of the half-supersymmetric  case,
are discussed in section \ref{ss:Vbranesd6}. In sections
\ref{ss:8ind6fromV5} and \ref{ss:BIhalfd4V3} we present the
half-supersymmetric DBI theories in $d=6$ and $d=4$ as they follow from
the V5 brane and V3 brane, respectively. Our conclusions are given in
section \ref{ss:discussion}. We have added three appendices. Appendix
\ref{app:A} contains the details of the calculation that yields the
maximal supersymmetric DBI action while Appendix \ref{app:B} contains a
similar calculation that leads to the half-supersymmetric DBI action.
Finally, some details on our notation are given in
Appendix~\ref{app:notation}.

\section{DBI-VA with Maximal 16+16 Supersymmetries in \texorpdfstring{$d=10$}{d=10} from the D9 superbrane}
 \label{ss:BIMaxd10D9}

\subsection{Bottom-Up From Supersymmetric Maxwell to Born-Infeld}

Our starting point is $\cN=1$, $d=10$ on-shell Maxwell described above
in (\ref{Maxwellaction}). It can be shown that requiring
$\epsilon$-supersymmetry up to order ${\cal O}(\alpha^2)$ for the
first-order non-derivative corrections of the on-shell multiplet leads
to the Born-Infeld combination \cite{Bergshoeff:1986jm}:\,
\footnote{With respect to \cite{Bergshoeff:1986jm} we have redefined
$\alpha^2 \rightarrow -\alpha^2/2$.}
\begin{eqnarray}
S &=& \int \rmd^{10}x\,\big\{-\tfrac{1}{4} F^2 + \bar\lambda \slashed{\partial} \lambda\big\} - 2\alpha c_4 F^{\mu\nu}\bar\lambda\Gamma_\mu\partial_\nu\lambda \nonumber\\[.2truecm]
&&+\tfrac{1}{8}\alpha^2\Big[\Tr F^4 -\tfrac{1}{4}\, \left(F^2\right)^2 +4 (1+4c_4^2) \big(F^2\big)^{\mu\nu}\bar\lambda\Gamma_\mu\partial_\nu\lambda\nonumber\\[.2truecm]
&& + (1-4c_4^2)F_\mu{}^\lambda\big(\partial_\lambda F_{\nu\rho}\big)\bar\lambda\Gamma^{\mu\nu\rho}\lambda + \ft{1}{2}(c_1+8c_4^2) F^2\bar\lambda \slashed{\partial}\lambda\nonumber\\[.2truecm]
&&-\ft{1}{2}c_2F_{\mu\nu}\big(\partial_\lambda F^\lambda{}_\rho\big)\bar\lambda \Gamma^{\mu\nu\rho}\lambda -\ft{1}{2} (c_3+4c_4^2)F_{\mu\nu}F_{\rho\sigma}\bar\lambda
\Gamma^{\mu\nu\rho\sigma} \slashed{\partial}\lambda \Big ]\nonumber\\
&&+{\cal O}(\alpha ^2\lambda ^4)+{\cal O}(\alpha ^3)\,.
\label{on-shellaction}
\end{eqnarray}
See Appendix \ref{app:notation} for our notations. The parameters $c_1,
c_2, c_3$ and $c_4$ cannot be determined. They are related to the
redefinitions
\begin{eqnarray}
A_\mu(0) &=& A_\mu - \tfrac{1}{16}\alpha^2 c_2 F^{\nu\rho}\bar\lambda\Gamma_{\mu\nu\rho}\lambda\,,\nonumber\\
\lambda(0) &=& \lambda + \tfrac{1}{2}\alpha c_4 F_{\mu\nu}\Gamma^{\mu\nu}\lambda +\tfrac{1}{32}\alpha^2c_1 F^2 \lambda
 -\tfrac{1}{32}\alpha^2c_3F_{\mu\nu}F_{\rho\sigma}\Gamma^{\mu\nu\rho\sigma}\lambda\,,
\label{redefinitionsci}
\end{eqnarray}
where $A_\mu(0)$ and $\lambda(0)$ are the fields for $c_i=0$.

The action (\ref{on-shellaction}) is invariant up to order $\alpha^2$ under the following supersymmetry transformations
\begin{eqnarray}
\label{on-shellcorrected1}
\delta_\epsilon  A_\mu &=& \bar\epsilon\Gamma_\mu\lambda +\tfrac{1}{2}\alpha c_4F^{\rho\sigma}\bar\epsilon\Gamma_\mu\Gamma_{\rho\sigma}\lambda\nonumber\\[.2truecm]
&&+\tfrac{1}{32}\alpha^2(c_1+2c_2-6)F^2\bar\epsilon\Gamma_\mu\lambda
+\tfrac{1}{8}\alpha^2(c_2-4)\big(F^2\big)_\mu{}^\nu\,\bar\epsilon\Gamma_\nu\lambda\nonumber\\[.2truecm]
&&-\tfrac{1}{16}\alpha^2(-c_2+2c_3+2)F_{\mu\nu}F_{\rho\sigma}\bar\epsilon\Gamma^{\nu\rho\sigma}\lambda\nonumber\\[.2truecm]
&&-\tfrac{1}{32}\alpha^2(c_2+c_3-1)F^{\rho\sigma}F^{\lambda\tau}\
\bar\epsilon\Gamma_{\mu\rho\sigma\lambda\tau}\lambda\nonumber\\
&&+{\cal O}(\alpha ^2\lambda ^3)+{\cal O}(\alpha ^3) \,.
\end{eqnarray}
\begin{eqnarray}
\delta_\epsilon \lambda &=& \tfrac{1}{4}\Gamma^{\mu\nu}F_{\mu\nu}\epsilon -\tfrac{1}{8}\alpha c_4\Gamma^{\mu\nu\rho\sigma}\epsilon F_{\mu\nu}F_{\rho\sigma}+\tfrac{1}{4}\alpha c_4 F^2\epsilon\nonumber\\[.2truecm]
&&-\tfrac{1}{128}\alpha^2(c_1+4c_3+48c_4^2-2)F^2 \Gamma^{ab}  F_{ab}\epsilon\nonumber\\[.2truecm]
&&-\tfrac{1}{16}\alpha^2(c_3+8c_4^2-1)\big(F^3\big)_{\mu\nu}\Gamma^{\mu\nu}\epsilon\nonumber\\[.2truecm]
&&+\tfrac{1}{384}\alpha^2(3c_3+24c_4^2+1)\Gamma^{\mu\nu\rho\sigma\lambda\tau}\epsilon F_{\mu\nu}F_{\rho\sigma}F_{\lambda\tau}\nonumber\\
&& -\alpha c_4\Gamma ^{\mu \nu }\lambda \bar \epsilon \Gamma _\nu \partial _\mu \lambda \nonumber\\
&&+{\cal O}(\alpha ^2\lambda ^2)+{\cal O}(\alpha ^3) \,.\label{on-shellcorrected2}
\end{eqnarray}
We note that there is no choice of redefinition parameters
$c_1,c_2,c_3,c_4$ possible such that the transformation rule of $A_\mu$
does not receive any $\alpha^2$-corrections. The order $\alpha $ terms
are all related to the lowest order using the redefinition with $c_4$ in
(\ref{redefinitionsci}).

The above result is valid in all dimensions where the  Majorana flip
relations are as in (\ref{flip}). For the higher fermion terms one needs
also the cyclic identity (\ref{Fierzid10}), restricting them to
$d=2,3,4,6$ and 10. Remark that when we choose
\begin{equation}
  c_1=-2\,,\qquad c_2=4\,,\qquad c_3=1\,,\qquad c_4=0\,.
 \label{cchoiced4}
\end{equation}
the transformation laws are not changed with respect to the lowest order
ones in $d=4$.

It appears difficult to continue to higher orders of deformation since
one has to find simultaneously the new terms in the supersymmetry
deformation and the new terms in the action deformation.

\subsection{A Complete Maximal Supersymmetric \texorpdfstring{$d=10$}{d=10} DBI-VA Model from the D9 superbrane}

The DBI model with complete set of 16+16 global supersymmetries is
defined by the gauge-fixed D9-brane action
\cite{Bergshoeff:1996tu,Aganagic:1996nn,Bergshoeff:1997kr,Kallosh:1997aw}
\begin{equation}
\label{Theaction}
S =  -\frac{1}{\alpha^2} \int \rmd^{10} x\,\left\{ \sqrt{- \det (G_{\mu\nu} + \alpha {\cal F}_{\mu\nu})}-1\right\} \,,
\end{equation}
where\footnote{Here we use the fact that the terms quartic in $\lambda$
in $b_{\mu\nu}$ vanish under the $\mu \leftrightarrow \nu$
anti-symmetrization, which allows a nice  covariant expression for the
2-form.}
\begin{eqnarray}
&&G_{\mu\nu} = \eta_{mn} \Pi_\mu^m \Pi_\nu^n \ , \qquad
\Pi_\mu^m = \delta_\mu^m - \alpha^2\bar\lambda \Gamma^m \partial_\mu \lambda \,, \qquad \mu=0,1,...,9\,, \qquad m=0,1,...,9\,,\nonumber\\
&&{\cal F}_{\mu\nu} \equiv F_{\mu\nu} -  b_{\mu\nu} \ , \qquad  b_{\mu\nu} = -\alpha \bar{\lambda}\Gamma_{m}\partial_{\mu}\lambda\, \Pi_{\nu}^{m}
-\left( \mu\leftrightarrow \nu \right)=-2\alpha \bar \lambda \Gamma _{[\nu }\partial _{\mu ]}\lambda \, .
\label{defGPicFD9}
\end{eqnarray}
The DBI action (\ref{Theaction}) has 16+16 global  supersymmetry
transformations, which are given in eqs. (89)-(92) in
\cite{Aganagic:1996nn} where the term in the transformations that is
non-linear in the vectors is somewhat complicated. Here we present an
explicit and relatively simple form of all these supersymmetries, based
also on \cite{Bergshoeff:1996tu,Bergshoeff:1997kr,Kallosh:1997aw}. The
detailed derivation starting from the $\kappa$-symmetric Dp-brane
actions is presented in Appendix \ref{app:A}. In short, we find from
(\ref{decompositionsgfeps}) and (\ref{decompositionsgfzeta}) the
following 16+16 supersymmetries:

\paragraph{Sixteen $\epsilon$ transformations, deformation of the Maxwell supermultiplet supersymmetries}

\begin{eqnarray}
\delta_{\epsilon} \lambda &=& -\tfrac{1}{2\alpha}\left( \mathbb{1}-
\betaplusprime\right)\epsilon
+\xi^{\mu}_{\epsilon}\partial_{\mu}\lambda\,, \nonumber \\
\delta_{\epsilon} A_\mu &=& -\tfrac{1}{2}\bar\lambda \Gamma_\mu\big(\mathbb{1} +
\betaplusprime\big)\epsilon
+\tfrac{1}{2} \alpha^2 \bar\lambda\Gamma_m (\tfrac{1}{3} \mathbb{1} +
\betaplusprime)
\epsilon \bar\lambda \Gamma^m \partial_\mu \lambda
+\xi^\rho_\epsilon F_{\rho\mu}\nonumber\\
&=& \alpha^{-1}\xi _{\epsilon \mu }+\xi^\rho  _\epsilon {\cal F}_{\rho \mu }-\alpha \xi _\epsilon ^\rho \bar \lambda \Gamma _\mu \partial _\rho \lambda
  + \ft13\alpha ^2\bar \epsilon \Gamma^m\lambda
  \bar \lambda \Gamma _m\partial ^\mu \lambda  \,, \label{epsilontrans}
\end{eqnarray}
\paragraph{Sixteen VA-type $\zeta$ transformations}
\begin{eqnarray}
\delta_{\zeta} \lambda &=& \alpha^{-1}\zeta +\xi ^\mu _\zeta\partial_{\mu}\lambda\,, \nonumber \\
 \delta_{\zeta} A_\mu&=&\alpha^{-1}\xi _{\zeta \mu }+\xi ^\rho  _\zeta F_{\rho\mu}
- \tfrac{1}{3} \alpha^2 \bar\lambda
\Gamma_m \zeta\bar\lambda \Gamma^m \partial_\mu \lambda \,. \label{decompositions}
\end{eqnarray}
where
\begin{equation}
\xi^{\mu}_{\epsilon} \equiv
-\tfrac{1}{2}\alpha\bar{\lambda}\Gamma^{\mu}\left(\mathbb{1} +
\betaplusprime\right)\epsilon\,,\qquad \xi ^\mu _\zeta = \alpha\bar\lambda \Gamma ^\mu \zeta \,,
\label{ximueps}
\end{equation}
and
\begin{equation}
  \betaplusprime=\cG\sum^{5}_{k=0} \frac{\alpha^k}{2^k k!} \hat \Gamma^{\mu_1
\nu_1 \cdots \mu_k \nu_k }\mathcal{F}_{\mu_1 \nu_1}\cdots
\mathcal{F}_{\mu_k \nu_k}\,,
 \label{betaplusprimeD9}
\end{equation}
where $\cG$ is defined in (\ref{defGF}). The expressions (\ref{ximueps})
are the quantities obtained in (\ref{xiepszeta}).

\subsection{Comparing Bottom-Up with Top-Down}
 \label{ss:comparing}

To obtain agreement between the step by step deformation of the Maxwell
theory with the complete Born-Infeld supersymmetric model
 following from the brane analysis we must choose our redefinition parameters in the bottom up approach such that
all bilinear fermion terms in the action that have a $\gamma^{(3)}$ or
higher-gamma structure vanish. Furthermore, by comparing the
supersymmetry rules with the brane answer, we deduce that the
$\big(F^3\big)_{\mu\nu}\gamma^{\mu\nu}$ structure in $\delta\lambda$
should be absent. Fitting the other structures that follow from the
brane analysis we find that there is a unique solution for the
redefinition parameters that gives agreement with the brane analysis.
This choice is given by
\begin{equation}
c_1=2\,,\qquad  c_2=0\,,\qquad  c_3=-1\,,\qquad  c_4=-\tfrac{1}{2}\,.
\label{preferredci}
\end{equation}
In this parametrization, the action is given by
\begin{eqnarray}
S&=&\int {\rm d}^{10} x\, \left\{ -\ft{1}{4} F^2 + \bar\lambda \slashed{\partial} \lambda+\alpha F^{\mu\nu}\bar\lambda\Gamma_\mu\partial_\nu\lambda\phantom{\big(F^2\big)^{\mu\nu}} \right.\nonumber\\
&&\left.+\ft{1}{8}\alpha^2\left[\Tr F^4-\ft{1}{4}\, \big(\Tr F^2\big)^2+8\big(F^2\big)^{\mu\nu}\bar\lambda\Gamma_\mu\partial_\nu\lambda
+ 2F^2\bar\lambda \slashed{\partial}\lambda\right]\right\} \nonumber\\
&&+{\cal O}(\alpha ^2\lambda ^4)+{\cal O}(\alpha ^3)\,.
\label{actiondownup}
\end{eqnarray}

Comparing this result with the brane analysis we may also deduce what
the next order corrections to the hidden $\zeta$-supersymmetry are. We
start with another basis of the transformations, using the shift
symmetry (\ref{eta}). The parameter $\eta $ in (\ref{eta}) refers to
another basis of the transformations. When we use as independent
parameters $\epsilon $ and $\eta $, the $\zeta$ transformations are
given by
\begin{equation}
  \delta_\zeta = \delta_\epsilon +\delta_\eta \,,\qquad \mbox{with}\qquad \epsilon =-\zeta \,,\qquad \eta =-2\zeta \,.
 \label{redefzetaeta}
\end{equation}

The $\eta $ transformations are for arbitrary coefficients $c_i$ given
by
\begin{eqnarray}
\delta_\eta  A^\mu&=&\ft{\alpha}{4 }\bar \eta F^{\nu\mu }\Gamma_\nu \lambda+\ft{\alpha}{8 }\bar \eta \Gamma^{\mu\nu\rho} F_{\nu\rho }\lambda
 -\ft{1}{16}\alpha c_2 F_{\nu\rho}\bar\eta\Gamma^{\mu\nu\rho}\lambda +{\cal O}(\alpha \eta \lambda ^3)+{\cal O}(\alpha ^2)
\,,\nonumber\\
  \delta_\eta \lambda&=& -\ft{1}{2\alpha }\eta+\alpha\left[-\ft{1}{32} F^2-\ft{1}{64}\Gamma^{\mu\nu\rho\sigma} F_{\mu\nu}F_{\rho\sigma} \right] \eta\nonumber\\
&&+\ft{1}{4}c_4 F_{\mu\nu}\Gamma^{\mu\nu}
\left[ \eta -\ft{1}{2}\alpha c_4 F_{\rho \sigma }\Gamma^{\rho \sigma }\eta \right]
\nonumber\\
&&+\ft{1}{64}\alpha c_1 F^2 \eta
 -\ft{1}{64}\alpha c_3F_{\mu\nu}F_{\rho\sigma}\Gamma^{\mu\nu\rho\sigma}\eta+{\cal O}(\alpha \eta \lambda ^2)+{\cal O}(\alpha ^2)
 \label{deletaalphacarb}
\end{eqnarray}
When using the parameters of (\ref{preferredci}) the $\epsilon $ transformations simplify to
\begin{eqnarray}
\delta_\epsilon  A_\mu &=& \bar\epsilon\Gamma_\mu\lambda -\ft{1}{4}\alpha \bar\epsilon\Gamma_\mu\Gamma\cdot F\lambda\nonumber\\[.2truecm]
&&-\ft{1}{8}\alpha^2F^2\bar\epsilon\Gamma_\mu\lambda
-\ft{1}{2}\alpha^2\big(F^2\big)_\mu{}^\nu\,\bar\epsilon\Gamma_\nu\lambda+\ft{1}{16}\alpha^2F^{\rho\sigma}F^{\lambda\tau}\
\bar\epsilon\Gamma_{\mu\rho\sigma\lambda\tau}\lambda\nonumber\\[.2truecm]
&&+{\cal O}(\alpha ^2\lambda ^3)+{\cal O}(\alpha ^3)\,,\nonumber\\
\delta_\epsilon \lambda &=& \ft{1}{4}\Gamma^{\mu\nu}F_{\mu\nu}\epsilon
+\ft{1}{16}\alpha \Gamma^{\mu\nu\rho\sigma}\epsilon
F_{\mu\nu}F_{\rho\sigma}
-\ft{1}{8}\alpha  F^2\epsilon\nonumber\\[.2truecm]
&&-\ft{1}{16}\alpha^2 F^2 \Gamma^{ab}  F_{ab}\epsilon+\ft{1}{96}\alpha^2\Gamma^{\mu\nu\rho\sigma\lambda\tau}\epsilon F_{\mu\nu}F_{\rho\sigma}F_{\lambda\tau}\nonumber\\[.2truecm]
&& +\ft12\alpha \Gamma ^{\mu \nu }\lambda \bar \epsilon \Gamma _\nu \partial _\mu \lambda \nonumber\\
&&+{\cal O}(\alpha ^2\lambda ^2)+{\cal O}(\alpha ^3)\,.\label{cupdowneps}
\end{eqnarray}
The $\eta $ transformations are then
\begin{eqnarray}
\delta_\eta  A_\mu &=&\alpha\ft{1}{8 }\bar \eta \Gamma \cdot F\Gamma _\mu \lambda
+{\cal O}(\alpha \eta \lambda ^3)+{\cal O}(\alpha ^2)
\,,\nonumber\\
  \delta_\eta \lambda&=& -\ft{1}{2\alpha }\eta-\ft{1}{8} \Gamma\cdot F
\left[ \eta +\ft{1}{4}\alpha  F\cdot \Gamma \eta \right]+{\cal O}(\alpha \eta \lambda ^2)+{\cal O}(\alpha ^2)
 \label{deletaalphacarbcup}
\end{eqnarray}

Using  (\ref{redefzetaeta}) one can verify that
\begin{eqnarray}
 \delta _\zeta A_\mu  & = & a_\mu(\zeta ) +\alpha a^\rho(\zeta ) F_{\rho \mu }
+{\cal O}(\alpha \eta \lambda ^3)+{\cal O}(\alpha ^2) \,,\nonumber\\
\delta _\zeta\lambda  & = & \alpha ^{-1}\zeta +\alpha a^\mu(\zeta )
\partial _\mu \lambda \,,\nonumber\\
&&  a^\mu(\zeta )\equiv \bar\lambda \Gamma ^\mu \zeta =-\bar \zeta\Gamma
^\mu \lambda\,.
 \label{decompositionszeta}
\end{eqnarray}

To expand the complete action (\ref{Theaction}) in orders of $\alpha $
we use the Mercator formula
\begin{eqnarray}
\det\left(\mathbb{1}+A\right) &=& \sum^{\infty}_{k=0}\frac{1}{k!}\left(-\sum^{\infty}_{j=1}\frac{\left(-1\right)^j}{j} \Tr\left(A^j\right)\right)^{k}\label{Mercator}\\
&=&1 + \Tr A
- \ft{1}{2}\Tr A^2 + \ft{1}{2} (\Tr A)^2 +\ft13 \Tr A^3-\ft12(\Tr A)(\Tr A^2)+\ft16(\Tr A)^3
+ \mathcal{O}\left(A^4\right)
\nonumber
\end{eqnarray}
where $A$ is an arbitrary dimensionless $n\times n$ matrix. To apply
this to (\ref{defGPicFD9}) we use
\begin{eqnarray}
 A_\mu {}^\nu &=& \alpha {\cal F}_{\mu }{}^\nu -\alpha ^2\bar \lambda \Gamma _\mu \partial ^\nu \lambda-\alpha ^2\bar \lambda \Gamma ^\nu \partial_\mu \lambda +
  \alpha ^4 (\bar \lambda \Gamma ^m\partial _\mu \lambda )(\bar \lambda \Gamma ^m\partial ^\nu \lambda )\nonumber\\
  &=& \alpha F_{\mu }{}^\nu -2\alpha ^2\bar \lambda \Gamma _\mu \partial ^\nu \lambda +
  \alpha ^4 (\bar \lambda \Gamma ^m\partial _\mu \lambda )(\bar \lambda \Gamma ^m\partial ^\nu \lambda )\,.
 \label{Aforactiondet}
\end{eqnarray}
The result agrees with (\ref{actiondownup}) and gives us moreover the
terms ${\cal O}(\alpha ^2\lambda ^4)$:
\begin{eqnarray}
S&=&\int {\rm d}^{10} x\, \left\{ - \ft14(F^2)+\bar \lambda \slashed{\partial} \lambda -\alpha F_{\mu }{}^\rho \bar \lambda \Gamma _\rho  \partial ^\mu \lambda
\right.
\nonumber\\
&&+\alpha ^2
   \left[ \ft18 (\Tr F^4)-\ft1{32} (F^2)^2+(F^2)_{\mu }{}^\rho \bar \lambda \Gamma _\rho  \partial ^\mu \lambda+
   \ft14\bar \lambda \slashed{\partial} \lambda(F^2)\right.\nonumber\\
   &&\left.\left.-\ft12(\bar \lambda \slashed{\partial} \lambda)^2 -\ft{1}{2} (\bar \lambda \Gamma ^m\partial _\mu \lambda )(\bar \lambda \Gamma ^m\partial ^\mu \lambda )
+\bar \lambda \Gamma _\mu \partial ^\rho  \lambda\bar \lambda \Gamma
_\rho  \partial ^\mu \lambda
   \right]+{\cal O}\left(\alpha ^3\right)\right\} \,
\label{actiondownupal2}
\end{eqnarray}
An easy check of the new terms is provided by calculating the variation
of the action under $\zeta$ symmetry proportional to $\alpha \lambda
^3$. These terms are not influenced by the redefinitions
(\ref{redefinitionsci}), and thus these can also be inserted in the
general expression (\ref{on-shellaction}), and thus also in
(\ref{actiondownup}). Other reparametrizations of the type $\lambda
\rightarrow c_5\alpha^2\lambda^2\partial\lambda$ do modify these terms.

The $\zeta $ supersymmetry rules (\ref{decompositions}) can be easily
expanded and agree with  (\ref{decompositionszeta}). To compare the
$\epsilon $ transformation rules, we need the following expansion of
(\ref{betaplusprimeD9}):
\begin{eqnarray}
 \betaplusprime&=& \cG\left[  1+\ft12\alpha \Gamma\cdot {\cal F}
 +\ft18\alpha ^2 \Gamma ^{\mu \nu \rho \sigma }F_{\mu \nu }F_{\rho \sigma }\right.\nonumber\\
 &&\left.+\ft12\alpha ^3\Gamma ^{\mu \nu \rho \sigma }F_{\mu \nu }\left(\bar \lambda \Gamma _{\sigma  }\partial _{\rho  }\lambda\right)
 +\ft1{48}\alpha ^3\Gamma ^{\mu \nu \rho \sigma \lambda \tau }F_{\mu \nu }F_{\rho \sigma }F_{\lambda \tau }
 +\alpha^3\Gamma^{\mu \rho}(\bar{\lambda}\Gamma^{\nu}\partial_{\rho}\lambda) F_{\mu \nu }\right] \nonumber\\
&& +{\cal O}(\alpha ^4)\,,\nonumber\\
 \cG&=& 1-\ft14\alpha ^2 F^2+\alpha ^3 F_{\mu \nu }\bar \lambda \Gamma^{\mu }  \partial^\nu \lambda
+{\cal O}(\alpha ^4)\,.
\end{eqnarray}
The first expression was again obtained using (\ref{Mercator}), where
now
\begin{eqnarray}
   A_\mu {}^\nu &=&\alpha  {\cal F}_{\mu \rho }G^{\rho \nu }=\alpha \left( F_{\mu \rho  }+ 2\alpha \bar \lambda \Gamma _{[\rho }\partial _{\mu ]}\lambda\right)
  \left( \eta ^{\rho \nu }+2\alpha ^2\bar \lambda \Gamma ^{(\rho  }\partial ^{\nu )}\lambda +{\cal O}(\alpha ^4)\right)\nonumber\\
  &=&\alpha F_{\mu}{}^\nu+ \alpha^2 \bar \lambda \Gamma ^\nu \partial _{\mu }\lambda
  - \alpha^2 \bar \lambda \Gamma_{\mu }  \partial^\nu \lambda +2\alpha ^3F_{\mu \rho  }\bar \lambda \Gamma ^{(\rho  }\partial ^{\nu )}\lambda +{\cal O}(\alpha ^4)\,,
  \label{AforcG}
\end{eqnarray}
such that traces of odd powers of $A$ vanish. In this way one obtains
\begin{eqnarray}
  \delta _\epsilon A_\mu &=& -\bar \lambda \Gamma_\mu \epsilon
   -\ft{1}{4}\alpha \bar\epsilon\Gamma_\mu\Gamma\cdot {\cal F}\lambda\nonumber\\[.2truecm]
&&-\ft{1}{8}\alpha^2F^2\bar\epsilon\Gamma_\mu\lambda
-\ft{1}{2}\alpha^2\big(F^2\big)_\mu{}^\nu\,\bar\epsilon\Gamma_\nu\lambda\nonumber\\[.2truecm]
&&+\ft{1}{16}\alpha^2F^{\rho\sigma}F^{\lambda\tau}\
\bar\epsilon\Gamma_{\mu\rho\sigma\lambda\tau}\lambda\nonumber\\[.2truecm]
&&+\alpha^2 \bar{\lambda}\Gamma^{\rho } \epsilon \bar \lambda (\Gamma _\mu \partial _\rho \lambda
  -\ft13\Gamma _\rho \partial _\mu \lambda)+{\cal O}(\alpha ^3)\,.
 \label{expanddelepsAmu}
\end{eqnarray}
This agrees with the bottom up results, adding the last line, and the
term implicit in the order $\alpha $ term with ${\cal F}$, as ${\cal
O}(\alpha ^2\lambda^3)$ terms. For the $\epsilon $ transformation of
$\lambda $ we find
\begin{eqnarray}
  \delta _\epsilon \lambda &=& \ft14\hat{\Gamma }\cdot {\cal F}\epsilon -\ft18\alpha\epsilon  {\cal F}^2
  +\ft{1}{16}\alpha \Gamma^{\mu\nu\rho\sigma}\epsilon
{\cal F}_{\mu\nu}{\cal F}_{\rho\sigma}\nonumber\\
&&-\ft{1}{16}\alpha^2 F^2 \Gamma^{ab}  F_{ab}\epsilon+\ft{1}{96}\alpha^2\Gamma^{\mu\nu\rho\sigma\lambda\tau}\epsilon F_{\mu\nu}F_{\rho\sigma}F_{\lambda\tau}\nonumber\\
&& -\alpha \partial _\mu \lambda \bar \lambda \Gamma ^\mu \epsilon -\ft14\alpha ^2\partial _\mu \lambda \bar \lambda \Gamma ^\mu
\Gamma \cdot F\epsilon+{\cal O}(\alpha ^3)\,.
 \label{expanddelepslambda}
\end{eqnarray}
Thus, there are higher order fermions included in the first term in
\begin{equation}
  \hat{\Gamma }^{\mu \nu }=\Gamma ^{\mu \nu }-2\alpha^2\Gamma^{\rho[\mu
}(\bar{\lambda}\Gamma^{\nu]}\partial_{\rho}\lambda)+{\cal O}(\alpha ^4)\,,
 \label{hatGammaexpand}
\end{equation}
in ${\cal F}$, and in the last term in (\ref{expanddelepslambda}). We
have agreement with (\ref{cupdowneps}) apart from the order $\alpha $
cubic fermion terms. It can be shown (after using the Fierz identity
(\ref{Fierzid10})) that the difference is proportional to a field
equation and is thus a `zilch symmetry'.\footnote{Note that also in the
comparison between different formulations of the VA actions
\cite{Volkov:1973ix,Casalbuoni:1988xh,Komargodski:2009rz} in
\cite{Kuzenko:2011tj}, such symmetries, also called `trivial
symmetries', were involved.}

\subsection{16+16 Supersymmetry Algebra in \texorpdfstring{$d=10$}{d=10}}

The algebra of the $\epsilon $ and $\zeta $ supersymmetries is
\begin{eqnarray}
[\delta(\epsilon_1), \delta(\epsilon_2)] &=& \delta_{cP}(\xi_{\epsilon \epsilon }^\mu)+\mbox{field equations} +{\cal O}(\alpha ^2)
\,,\qquad
\xi_{\epsilon \epsilon }^\mu = {\bar\epsilon}_1\Gamma^\mu\epsilon_2\,,
\nonumber\\[.2truecm]
[\delta(\epsilon), \delta(\zeta)] &=& \delta_{cP}(\xi _{\epsilon \zeta  }^\mu )+\delta _{U(1)}(\Lambda_{\epsilon \zeta } )\,, \qquad \xi _{\epsilon \zeta  }^\mu =\bar{\zeta}\Gamma^{\mu }\epsilon\,,\nonumber\\
&& \Lambda_{\epsilon \zeta}= x_\mu \alpha^{-1}\xi _{\epsilon \zeta  }^\mu+ \ft13\xi _\zeta ^m\bar \lambda \Gamma _m\betaplusprime \epsilon
+\xi _\zeta ^\rho  \xi _\epsilon ^\sigma \bigl(F_{\rho\sigma}-\ft{1}{3}\alpha\bar{\lambda}\Gamma_{\rho}\partial_{\sigma}\lambda+\alpha\bar{\lambda}\Gamma_{\sigma}\partial_{\rho}\lambda\bigr) \nonumber \\
&&
+\ft{1}{3}\alpha^2\xi _\zeta ^\rho (\bar\lambda\Gamma^m{\epsilon})(\bar{\lambda}\Gamma_m\partial_{\rho}\lambda)\,,\nonumber\\[.2truecm]
[\delta(\zeta_1), \delta(\zeta_2)] &=&\delta_{cP}(\xi _{\zeta  \zeta  }^\mu )+\delta _{U(1)}(\Lambda_{\zeta  \zeta  } )\,, \qquad \xi _{\zeta  \zeta  }^\mu =2\bar \zeta _1\Gamma^\mu \zeta _2\,,\nonumber\\
&&\Lambda_{\zeta \zeta }=\frac{1}{\alpha }\xi _{\zeta  \zeta  } ^\mu x_\mu + \xi _{\zeta 2}^\rho \xi _{\zeta 1} ^\sigma \left( F_{\rho \sigma }-\ft23\alpha \bar \lambda \Gamma _{[\rho} \partial _{\sigma]} \lambda\right)\,,
\end{eqnarray}
where $\xi _\epsilon $ and $\xi _\zeta $ are the expressions in
(\ref{ximueps}).
 Here $\delta_{cP}$ is a covariant translation, i.e. a spacetime
translation combined with an Abelian gauge transformation $\delta A_\mu
=
\partial_\mu\Lambda$ with parameter $\Lambda=-\xi^\mu  A_\mu $.
The first commutator is only valid on-shell. When using to this order
the transformations of $\lambda $ as obtained in (\ref{cupdowneps}) the
commutator closes without using field equations on $A_\mu $, and using
field equations on the fermion field. However, with the transformations
as in (\ref{expanddelepslambda}), the field equations are also needed
for the commutator on the gauge field.

Note that the first terms of $\Lambda_{\epsilon \zeta} $ and
$\Lambda_{\zeta \zeta }$ (proportional to $x^\mu $) can also be
understood as shift transformations of the vector field.

\section{DBI-VA with Maximal 16 + 16  Supersymmetry in \texorpdfstring{$d=4$}{d=4} from the D3 superbrane}
\label{ss:16d4fromD3}
The complete DBI action in $d=4$  is relatively simple when the fermion
part of the action  is ``packaged into a $d=10$ form'', namely all
spinors are still $d=10$ Majorana-Weyl spinors and all $\Gamma^{m'}$ and
$\Gamma^I$ are $d=10$ matrices. We find
\begin{equation}
\label{4actionBI} S =  -\frac{1}{\alpha^2} \int \rmd^{4} x\,\left\{
\sqrt{- \det (G_{\mu\nu} + \alpha {\cal F}_{\mu\nu})} - 1 \right\}\,, \qquad \mu=0,1,2,3\,,
\end{equation}
where
\begin{eqnarray}
G_{\mu\nu} &=& \eta_{m n} \Pi_\mu^{m}
\Pi_\nu^{n}= \eta_{m' n'} \Pi_\mu^{m'}
\Pi_\nu^{n'}+\delta_{IJ} \Pi_\mu^I\Pi_\nu^J \,, \qquad  m'=0,1,2,3 \,, \qquad I=1,...,6\,,\nonumber \\[.2truecm]
\Pi_\mu^{m'} &=& \delta^{m'}_{\mu}
-\alpha^2\bar\lambda \Gamma^{m'}
\partial_\mu \lambda \ , \qquad \Pi_\mu^I = \partial_\mu\phi^I
-\alpha^2\bar\lambda \Gamma^I
\partial_\mu \lambda\,, \qquad {\cal F}_{\mu\nu} \equiv F_{\mu\nu} - b_{\mu\nu} \,,\nonumber\\
 b_{\mu\nu} &=&2\alpha\bar{\lambda}\Gamma_{[\mu }\partial_{\nu]}\lambda
-2\alpha\bar{\lambda}\Gamma_{I}\partial_{[\mu}\lambda\partial_{\nu]}
\phi^I =-2\alpha\bar{\lambda}\Gamma_{m'}\partial_{[\mu}\lambda\, \Pi_{\nu]}^{m'}
-2\alpha\bar{\lambda}\Gamma_{I}\partial_{[\mu}\lambda \, \Pi_{\nu]}
^I  \, .
\label{defGPicFD3}
\end{eqnarray}
This action has a maximal number of  supersymmetries. The 16
$\epsilon$-supersymmetries correspond to a deformation of the original
16 supersymmetries of the $\cN=4$, $d=4$ Maxwell multiplet, while the 16
$\zeta$-supersymmetries correspond to VA-type supersymmetries.
Explicitly, we find the following transformation rules:


\paragraph{Sixteen $\epsilon$ transformations, deformation of the Maxwell supermultiplet}


\begin{eqnarray}
\delta_\epsilon  \phi^I &=& \ft12\alpha\bar\lambda\Gamma^I\left[\unity+\betaplusprime\right] \epsilon
+\xi^\mu_\epsilon \partial_\mu \phi^I\,, \nonumber \\
\delta_\epsilon  \lambda &=&- \ft1{2\alpha}\left[\unity  -
\betaplusprime\right]\epsilon
+\xi^{\mu}_\epsilon \partial_{\mu}\lambda\,, \nonumber \\
 \delta_\epsilon  A_\mu&=&-\ft12\bar \lambda\big(
 \Gamma_\mu + \Gamma_I\partial_\mu\phi^I\big) \left[\unity + \betaplusprime\right]\epsilon
 \nonumber \\
&&+\ft12 \alpha^2 \bar \lambda \Gamma_m\left[ \tfrac{1}{3}\unity  +\betaplusprime\right] \epsilon
 \bar\lambda \Gamma^m \partial_\mu \lambda+\xi_\epsilon ^\rho F_{\rho\mu} \,.
 \label{epsilontransMaxw}
\end{eqnarray}

\paragraph{Sixteen VA-type $\zeta$ supersymmetry transformations}

\begin{eqnarray}
\delta_{\zeta} \phi^I &=& -\alpha{\bar\lambda}\Gamma^I\zeta
+\xi ^{\mu}_\zeta \partial_\mu \phi^I\,, \nonumber \\
\delta_{\zeta} \lambda &=& \alpha^{-1}\zeta +\xi ^{\mu}_\zeta\partial_{\mu}\lambda\,, \nonumber \\
 \delta_{\zeta} A_\mu&=&\bar\lambda \big(
 \Gamma_\mu + \Gamma_I\partial_\mu\phi^I\big)\zeta
+ \xi ^{\rho}_\zeta F_{\rho\mu}
-\tfrac{1}{3} \alpha^2 \bar\lambda
\Gamma_m \zeta \bar\lambda \Gamma^m \partial_\mu \lambda \,,
\label{decompositions1}
\end{eqnarray}
where $\xi^{\mu}_{\epsilon}$ and $\xi _\zeta ^\mu $ are given in
(\ref{ximueps}),
\begin{eqnarray}
    \betaplusprime&=&-\rmi\cG\sum^{2}_{k=0} \frac{\alpha^k}{2^k k!} \hat \Gamma^{\mu_1
\nu_1 \cdots \mu_k \nu_k }\mathcal{F}_{\mu_1 \nu_1}\cdots
\mathcal{F}_{\mu_k \nu_k}\Gamma _{(0)}^{D3}\Gamma _*^{(3)}=1+{\cal O}(\alpha )\,,\nonumber\\
  \Gamma_{(0)}^{D3} &=&
\frac{1}{4!\sqrt{|G|}}\varepsilon^{\mu_1\dots \mu_{4}}\hat \Gamma_{\mu_1\dots
\mu_{4}}=\rmi\Gamma _*^{(3)}+{\cal O}(\alpha )\,,\qquad\Gamma _*^{(3)}=-\rmi\Gamma ^0\Gamma ^1\Gamma ^2\Gamma ^3\,,
 \label{betaplusprimeD3}
\end{eqnarray}
and terms with $\partial _\mu \phi ^I$ are considered as ${\cal
O}(\alpha )$.

The action also has a shift symmetry
\begin{equation}
\delta\phi^I= a^I\,.
\label{shift}
\end{equation}
The scalars in the $d=4$ action (\ref{4actionBI})  originate from the 6
directions $X^I$ transverse to the D3-brane, see Appendix \ref{app:A}.
This symmetry is a surviving part of the Poincar{\'e} translation symmetry
in $d=10$, $X^{I}\rightarrow X^I+ a^I$ for $I=1,...,6$.

\subsection{16+16 Supersymmetry Algebra in \texorpdfstring{$d=4$}{d=4}}

One can use $d=10$ algebra via dimensional reduction to  $d=4$ or work
out the explicit $d=4$ algebra directly. Here we are mostly interested
in the way the second supersymmetry and the combination of the first and
second acts on the scalars.
\begin{eqnarray}
[\delta(\zeta_1),\delta(\zeta_2)]\phi^I &=& 2\bar{\zeta}_2(\Gamma^I-\Gamma^{\mu}\partial_{\mu}\phi^I)\zeta_1\,,\nonumber\\[0.2truecm]
[\delta(\epsilon),\delta(\zeta)]\phi^I &=& \bar{\zeta}(\Gamma^{\mu}\partial_{\mu}\phi^I-\Gamma^I)\epsilon\,.
\label{shiftcomm}
\end{eqnarray}
A shift symmetry on scalars is an  important feature of duality
symmetries: in particular, in the case of $E_{7(7)}$ symmetry in $N=8$
supergravity  the single-soft scalar limits were studied in detail in
\cite{Beisert:2010jx} and used to prove the UV finiteness below 7 loop
order.
Thus we see from (\ref{shiftcomm}) that the shift of scalars that is a
symmetry in the action appears already in the algebra of the extra
supersymmetries.

\subsection{Maxwell \texorpdfstring{$d=4$}{d=4} Supermultiplet with \texorpdfstring{$SU(4)$}{SU(4)} symmetry}
\label{ss:Maxwd4N4}

We would like to rewrite the fermionic sector of the action
(\ref{4actionBI}) and its symmetries using the four $d=4$ Majorana
fermions and the $d=4$ $\gamma$-matrices, following
\cite{Brink:1976bc,Gliozzi:1976qd}  where the dimensional reduction from
10 to 4 was performed in SYM theory, and the  $SU(4)$ symmetry of the
model was revealed. Moreover, we would like to bring the complete action
to the form which at the quadratic level coincides with the one in
\cite{de Roo:1984gd,Ferrara:2012ui} where we have a $d=4$, $\cN=4$
Maxwell multiplet. We would like to deform the $\cN=4$ Maxwell action,
namely
\begin{equation}
S_{\rm Maxw}= \int \rmd^4x \Big (-\ft14 F_{\mu\nu} F^{\mu\nu }  + 2\bar
\psi_i \slashed{\partial} \psi^i -\ft18\partial _\mu\varphi_{ij}
\partial^\mu \varphi^{ij}\Big ) \,,
\label{quadr}
\end{equation}
where in \cite{de Roo:1984gd} the $d=4$ left-handed chiral spinor is
assigned to the fundamental representation of $SU(4)$, and carries an
upper $SU(4)$ index; right-handed components will then transform
according to the conjugate representation, in agreement with the
Majorana property and have a lower index:
\begin{equation}
\psi_i=-\gamma_* \psi_i \, , \qquad \psi^i\equiv  (\psi_i)^C =\gamma_* \psi^i\, ,
\qquad \varphi^{ij} \equiv (\varphi_{ij})^*= -\ft12 \varepsilon^{ijk\ell } \varphi_{k\ell }\,.
\end{equation}
The action (\ref{quadr}) is invariant under the 16 supersymmetries
\begin{equation}
\delta_\epsilon  A_\mu= \bar \epsilon ^i \gamma_\mu \psi_i + \bar \epsilon _i  \gamma_\mu  \psi^i\, , \qquad \delta \psi_i= \ft14 \gamma\cdot F^+ \epsilon_i
+ \ft12\slashed{\partial} \varphi_{ij} \epsilon^j\, , \qquad \delta\varphi_{ij} = 4\bar \epsilon_{[i} \psi_{j]}-2\varepsilon_{ijk\ell } \bar \epsilon^k \psi^\ell\,.
\label{epsN4d4}
\end{equation}
The second, trivial, set of $\eta$ symmetries is
\begin{equation}
\delta A_\mu=0\, , \qquad \delta \varphi^{ij}=0\, , \qquad \delta \psi_i=  -\tfrac{1}{2\alpha }\eta_i\, , \qquad \delta \psi^i=  -\tfrac{1}{2\alpha }\eta^i\,.
\label{etatransfd4N4}
\end{equation}
We can write the $\varphi _{ij}$ in terms of 6 real components as
\begin{equation}
\varphi_{ij}\equiv  \varphi_a \beta_{ij}^a -{\rm i} \varphi_{a+3} \alpha ^a_{ij}\,
, \qquad \varphi^{ij} \equiv (\varphi_{ij})^*= -\ft12 \varepsilon^{ijkl}
\varphi_{kl}\,,
 \label{rule}
\end{equation}
where $a=1,2,3$ and $\alpha$, $\beta$ are the real antisymmetric
$4\times 4$ matrices of $SU(2)\times SU(2)$ given explicitly in
\cite{Gliozzi:1976qd} and whose properties are given in
(\ref{propalbe}).

Now we indicate how this can be seen as the $\alpha =0$ limit of the
action (\ref{4actionBI}) and the transformation rules
(\ref{epsilontransMaxw}). We therefore define the 6 scalars $\phi ^I$
and $d=10$ chiral Majorana spinor $\lambda $ as
\begin{equation}
  \phi ^I=\alpha \varphi^I\,,\qquad \lambda =\begin{pmatrix}\psi   ^i\cr\psi    _i\end{pmatrix}\,,\qquad
  \bar \lambda =\begin{pmatrix}\bar\psi   _i&\bar\psi ^i\end{pmatrix}\,,
 \label{philambda10to4}
\end{equation}
using the decomposition of the $d=10$ spinors in $d=4$ spinors according
to the Clifford representation (\ref{Gamma104}). Contractions of two
$d=10$ vectors in a $d=4$ part and the $SO(6)$ part is done according to
\begin{equation}
 A^mB_m=A^\mu B^\nu \eta_{\mu \nu} + A^I B^J \delta _{IJ}= A^\mu B^\nu \eta_{\mu \nu} + ( A^a B^b \delta _{ab} +  A^{a+3} B^{b+3} \delta _{ab})\,,
\end{equation}
and when using the scalars as in (\ref{rule}) this implies that
\begin{equation}
 \varphi _a \varphi _b \delta ^{ab} +  \varphi _{a+3} \varphi _{b+3} \delta ^{ab} = \ft1{8}\varphi _{ij} \varphi ^{ij}\,.
\label{translateIIij}
\end{equation}
Using also for the $\epsilon $ parameters the decomposition as in
(\ref{philambda10to4}) (and for convenience using $\epsilon^i$ and
$\epsilon _i$ as components of $\epsilon $) we can write e.g.
\begin{eqnarray}
 {\bar\epsilon}\Gamma_\mu\lambda  &=& \bar \epsilon_{i }\gamma_\mu \psi ^i +\bar \epsilon^i \gamma_\mu \psi _i \,,\nonumber\\
{\bar\epsilon}\Gamma^a\lambda  &=&-\bar \epsilon_i \beta ^{a\,ij}\psi  _j-\bar \epsilon^i \beta ^a_{ij}\psi  ^j\,,\nonumber\\
{\bar\epsilon}\Gamma^{a+3}\lambda  &=&-\rmi\bar \epsilon_i \alpha ^{a\,ij}\psi  _j+\rmi\bar \epsilon^i \alpha  ^a_{ij}\psi  ^j\,.
 \label{translatederivlambda}
\end{eqnarray}
This allows to check that the $\alpha =0$ part of the action
(\ref{4actionBI}) and transformations (\ref{epsilontransMaxw})agree with
the action (\ref{quadr}) and transformations (\ref{epsN4d4}). Also the
lowest order ($\alpha ^{-1}$) of (\ref{decompositions1}) equals
(\ref{etatransfd4N4}) using the translation (\ref{redefzetaeta}).
Therefore the full action and transformations (\ref{4actionBI}) and
(\ref{epsilontransMaxw}) are a deformation of the lowest order $d=4$,
$\cN=4$ theory.



The form of the action and supersymmetries in $SU(4)$, $d=4$ fermion
notation is significantly more complicated than the one above with
$d=10$ packaging of fermions in $d=4$ action and in supersymmetry rules.
One can view this fact as a matter of notational convenience. Using the
four four-component Majorana spinors in $d=4$  leads to an increasing
complexity of the complete non-linear action and its supersymmetries. We
therefore leave it in the form given in
(\ref{4actionBI})-(\ref{decompositions1}) with understanding that it
codifies all information which may, in principle, be expressed also
using the $d=4$ spinors and  $SU(4)$ symmetry of the theory, as in the
linearized action in (\ref{quadr}).

\section{Vector Superbranes in \texorpdfstring{$d=6$}{d=6} and DBI-VA dynamics}
\label{ss:Vbranesd6}



In the previous part of the paper and in the appendix \ref{app:A} we
have constructed the worldvolume action of Dirichlet branes in a flat
background with 32 supersymmetries. These D-branes occur as solutions of
IIA and IIB supergravity. A characteristic feature of these Dirichlet
branes is that their worldvolume dynamics is described by a vector
supermultiplet. The scalars in this multiplet are the embedding scalars
and the vector is the Born-Infeld vector. A special case is the D9-brane
which has no embedding scalars at all. We have seen how, starting from a
kappa-symmetric worldvolume action of the D9-brane, this has led, after
gauge-fixing, to a supersymmetric DBI action with 16+16 supersymmetries.
The existence of this D9-brane suggests the existence of all other
Dp-branes, with $0\le p\le 8$,  by dimensional reduction of the
world-volume. More precisely, reducing the $d=10$ supersymmetric DBI
action to $p+1$ dimensions leads to a supersymmetric DBI action in $p+1$
dimensions with $d-p-1$ scalars and one DBI vector. This is precisely
the worldvolume content of the Dp-brane.

In Appendix \ref{app:B} we describe in a similar way brane actions with
vector multiplets and 8+8 supersymmetries.
Such branes do not occur in $d=10$ Heterotic or Type I supergravity. In
the case of 9-branes, this is consistent with the fact that vector
multiplets with 8+8 supersymmetries only occur in $d\le 6$ dimensions.
It is therefore natural to look for branes with vector multiplets and
8+8 supersymmetries in $d=6$ half-maximal supergravity. There are three
half-maximal $d=6$ supergravities: Heterotic and  Type I, with
non-chiral (1,1) supersymmetry,  and chiral iib supergravity, with (2,0)
supersymmetry. Note that the Heterotic theory is S-dual to Type I and
that the Type I theory is T-dual to iib, see, e.g.,
\cite{Behrndt:2001ab}. The heterotic supergravity contains the (1,1)
supergravity with 20 vector multiplets. Apart from the scalars that
transform in the $SO(4,20)$ isometry group, it contains one `dilaton'
that is invariant under this group. The iib theory is a (2,0)
supergravity coupled to 21 tensor multiplets, whose scalars transform
under a $SO(5,21)$ isometry group. There is no other invariant scalar,
and therefore no corresponding string coupling constant.\footnote{Both
theories reduce to the $\cN=4$ theory in $d=5$, where the `dilaton' is
present}

In searching for branes with a worldvolume vector multiplet, it is
important to keep in mind that these branes are not necessarily
Dirichlet branes in the sense that their tensions scale with the inverse
string coupling constant. In fact, $d=6$ Heterotic supergravity has no
branes whose tension scales with $g_s^{-1}$, whereas the Type I theory
has only such branes with worldvolume hypermultiplets. For our purposes,
however, all we need is a brane whose worldvolume dynamics is described
by a vector multiplet. We do not mind that the tension of such branes
scale differently than the Dirichlet branes. Let us call from now the
branes with a worldvolume vector multiplet `vector branes'.

We focus here  on the space-filling vector 5-branes (called V5-branes)
with a 6-dimensional worldvolume and the vector 3-branes (V3-branes)
with a 4-dimensional worldvolume. Note that the  V5-branes couple to
6-forms, which are not part of the supergravity multiplet that describes
physical degrees of freedom. The V3-branes are `defect-branes' with two
transverse directions. They couple to 4-forms that are dual to the
scalars of the supergravity multiplet. In \cite{Bergshoeff:2012jb} an
analysis was given of the worldvolume content of the different branes of
$d=6$ half-maximal supergravity. This analysis was based on the
construction of a gauge-invariant Wess-Zumino (WZ) term consistent with
worldvolume supersymmetry. The requirement that such a WZ term can be
constructed is a necessary condition for a kappa-symmetric worldvolume
action to exist.
In this work we will only consider branes in a flat background. Based on
this WZ analysis it was concluded in \cite{Bergshoeff:2012jb} that
V5-branes occur both in $d=6$ Heterotic, Type I and iib supergravity
whereas  V3-branes only occur in $d=6$ iib supergravity. In all cases
the tension of the branes is not proportional to the inverse string
coupling constant like it was the case for the tension of the IIA and
IIB D-branes. For our purposes, it is sufficient to restrict to the
V-branes of $d=6$ iib supergravity.

Following the analysis of appendix \ref{app:A} we  construct in Appendix
\ref{app:B} the supersymmetric DBI actions with 8+8 supersymmetries
describing the Vp-branes ($p=1,3,5$) of $d=6$ iib supergravity. We use
these branes to find the half-maximal susy DBI actions.\footnote{Note
that  V1-branes are special in the sense that a vector on a
two-dimensional worldvolume is equivalent to an integration constant.}

\section{DBI-VA with Half-Maximal 8+8 Supersymmetry in \texorpdfstring{$d=6$}{d=6} from the V5 brane}
\label{ss:8ind6fromV5}

We could have used the D5 brane of IIB $d=10$ theory to get the DBI in
 $d=6$ with maximal supersymmetry and, of course, we could have truncated
it to the half of supersymmetry. Alternatively, we may use the vector
branes available in $d=6$ which have been constructed in Appendix
\ref{app:B}. The $d=6$ DBI action with 8+8 supersymmetries (using  V5
brane) is given by
\begin{equation}
S =  -\frac{1}{\alpha^2} \int \rmd^{6} x\,\left\{
\sqrt{- \det (G_{\mu\nu} + \alpha {\cal F}_{\mu\nu})} - 1 \right\}\ ,
\end{equation}
where $\mu=0,1,...,5$. The quantities that appear here are the same
expressions as in  (\ref{defGPicFD9}), where now $m=0,1,...,5$. Here
$\lambda$ is a symplectic $d=6$ Majorana-Weyl spinor and the symplectic
indices $i$ in a bilinear are contracted using the antisymmetric
$\varepsilon_{ij}$, see (\ref{symplecticcontraction}). The above DBI
action is invariant under the eight $\epsilon$ transformations,
deformation of the $d=6$ Maxwell  supersymmetries, and eight VA-type
$\zeta$ transformations as expressed in (\ref{epsilontrans}) and
(\ref{decompositions}). We use now for $\Gamma^m$ and $\Gamma_\mu$ the
flat $d=6$ matrices.
\begin{equation}
  \betaplusprime=\cG\sum^{3}_{k=0} \frac{\alpha^k}{2^k k!} \hat \Gamma^{\mu_1
\nu_1 \cdots \mu_k \nu_k }\mathcal{F}_{\mu_1 \nu_1}\cdots
\mathcal{F}_{\mu_k \nu_k}=1+{\cal O}(\alpha )\,,
 \label{betaplusprimeV5}
\end{equation}
based on the matrices $ \hat \Gamma_{\mu}$, which are the pull-back to
the word-volume matrix of $d=6$.

\section{DBI-VA with Half-Maximal 8 + 8 Supersymmetry in \texorpdfstring{$d=4$}{d=4} from the V3 brane}
\label{ss:BIhalfd4V3}

As in the maximal supersymmetry case, the $d=4$ DBI model with
half-maximal supersymmetry is relatively simple when all fermions are
packaged in $d=6$ symplectic Majorana fermions and all the
$\Gamma$-matrices are the ones in $d=6$.
\begin{equation}
S =  -\frac{1}{\alpha^2} \int \rmd^{4} x\,\left\{
\sqrt{- \det (G_{\mu\nu} + \alpha {\cal F}_{\mu\nu})} - 1 \right\}\,, \qquad \mu=0,1,2,3\,,
\end{equation}
where the definitions (\ref{defGPicFD3}) apply with now $m=0,\ldots ,5$,
$m'=0,\ldots 3$ and $I=1,2$. The 8+8 supersymmetries are given by
(\ref{epsilontransMaxw}), (\ref{decompositions1}) where, however, the
spinors are $d=6$ symplectic Majorana spinors and the $\Gamma$'s are
those from $d=6$.


Finally, we comment on the $d=4$ DBI action with half-maximal
supersymmetry and $SU(2)$ symmetry. If we would have a simple action for
$d=4$ DBI with maximal supersymmetry and $SU(4)$ symmetry, truncating it
to half-maximal case with $SU(2)$ symmetry would be extremely simple, we
would just allow the $SU(4)$ index take values not in $i=1,2,3,4$ but in
$i=1,2$. However, as explained above, the complete action is simple only
when fermions are in the higher dimensional form. So, here, the
procedure of switching to the DBI action with manifest $SU(2)$ would be
the same as in the previous case with manifest $SU(4)$. The expressions
for the action and supersymmetries become complicated, but it is clear
that in principle all information in the BI action above  can be
transferred into an $SU(2)$ covariant action.

\section{Discussion}
\label{ss:discussion}

We presented here an explicit  completion to all orders of the
deformation of the Maxwell supermultiplets with maximal supersymmetry in
 $d=10$, 4 and half-maximal ones in $d=6$, 4. The deformation of the global
supersymmetry of the Maxwell multiplet to all orders is required. It is
also accompanied by a non-linear extra supersymmetry of the
Volkov-Akulov type of the same dimension: the maximal case has in total
16+16 supersymmetries and the half-maximal one has in total 8+8 of these
supersymmetries. Both our maximal supersymmetry and half-maximal
supersymmetry models are realized in the DBI type actions: when all
spinors and scalars are absent, we recover the classical BI models, for
example in $d=10$ we find
\begin{equation}
\label{simpleBI}
S_{\rm BI} =  -\frac{1}{\alpha^2} \int \rmd^{10} x\,\left\{ \sqrt{- \det (\eta_{\mu\nu} + \alpha  F_{\mu\nu})}-1\right\} \ ,
\end{equation}
Also, it is interesting that in $d=10$, when the covariant 2-form ${\cal
F}_{\mu\nu}$ is absent, the same action is a $d=10$ analog, as discussed
in \cite{Kallosh:1997aw}, of the $d=4$ Volkov-Akulov action
\cite{Volkov:1973ix} (at $\alpha=1$ and ignoring the constant part of
the action)
\begin{equation}\label{simpleVA}
S_{\rm VA} =  - \int {\rm d}^{10} x\,
\sqrt{- \det G_{\mu\nu} } = \int E^{m_0} \wedge ...\wedge E^{m_9}
\end{equation}
\begin{equation}\label{E}
E^m={\rm d} x^m+ \bar\lambda \Gamma^{m}
{\rm d} \lambda \,.
\end{equation}
The general maximal (half-maximal) supersymmetric
Dirac-Born-Infeld-Volkov-Akulov models presented in this paper, are only
slightly more complicated than the DBI and VA actions shown above. The
Lagrangian is always of the form
\begin{equation}
S_{\rm DBI-VA} = -\frac{1}{\alpha^2}\Big (\sqrt{- \det (G_{\mu\nu} + \alpha {\cal F}_{\mu\nu})} - 1\Big)
\end{equation}
where $G_{\mu\nu}$ and ${\cal F}_{\mu\nu}$ are defined for each case for
$d=10$, $d=4$ in the maximum and $d=6$, $d=4$ in the half-maximum
supersymmetry, in the relevant sections of the paper. The all order in
deformation supersymmetries are in all cases given by rather complicated
expressions, which involves the 2-forms and the fermion dependent
pull-back world-volume matrices, related to (\ref{E}). However, the
exact hidden  non-linear supersymmetry transformation of fermions is a
simple shift and quadratic in fermions expression which is literally the
original Volkov-Akulov formula
\begin{equation}
\delta_{\zeta} \lambda = \alpha^{-1}\zeta +\alpha
\bar{\lambda}\Gamma^{\mu}\zeta \partial_{\mu}\lambda\,.
\end{equation}
Despite the complicated dependence of supersymmetries on ${\cal
F}_{\mu\nu}$, the complete models in $d=10$, 6, 4 are given on one page
each, using a notation which we called `fermions packaged in a $d=10$
($d=6$) form' for the maximal (half-maximal)  case.

It would be interesting to compare our new models with the known
half-maximal models in $d=4$
\cite{Bagger:1996wp,Rocek:1997hi,Tseytlin:1999dj,Ketov:1998ku,Ketov:2001dq,Kuzenko:2000tg,Kuzenko:2000uh,Bellucci:2001hd,Sorokin:1999jx,Pasti:2000zs,Carrasco:2011jv,Chemissany:2011yv,Broedel:2012gf}.
with various amount of supersymmetry. The models with 8+8
supersymmetries are known only up to a certain level of deformation,
whereas our model with the same amount of supersymmetries is known to
all orders of deformation.\footnote{The existence of the complete 8+8
DBI-VA  model in $d=4$ described in this paper helps to explain certain
puzzles concerning hidden supersymmetry and duality of some
4-dimensional models with manifest $\cN=2$ supersymmetry and hidden
$\cN=2$ supersymmetry \cite{Carrasco:2013qia}.} In  models studied in
\cite{Bellucci:2001hd,Carrasco:2011jv,Broedel:2012gf}, 8 supersymmetries
are manifest and undeformed, whereas the hidden 8 are deformed. The
comparison between these models and our complete model will likely be
possible with account of field redefinitions which relate various
versions of the VA models even in absence of the vector field. Recently
is was shown how various forms of a goldstino action are related to each
other by a non-linear local field redefinition
\cite{Casalbuoni:1988xh,Kuzenko:2005wh,Komargodski:2009rz,Liu:2010sk,Zheltukhin:2010xr,Kuzenko:2010ef,Kuzenko:2011tj}.

The DBI models with 16+16 supersymmetries do not seem to be given in the
literature. Part of the reason for the absence of such models is that in
case of 8+8 supersymmetries, the $\cN=2$, $d=4$ superfields were used.
However for $\cN=4$ no off shell superfields are available, therefore
the tools used for 8+8 models are not available for 16+16. Meanwhile we
have shown that using the superbrane approach it is not necessary to
keep any linear supersymmetries manifest. The theory includes a  natural
Born-Infeld non-supersymmetric model and, in absence of 2-form ${\cal
F}_{\mu\nu}$, has  natural Goldstino variables in which the action has
an original Volkov-Akulov form.

Finally, we would like to discuss our original goal to find new ways  to
construct supersymmetric invariants for theories with extended
supersymmetries where there are no known auxiliary fields. We have
learned here how to build the deformation of the $\cN=4$, $d=4$ Maxwell
multiplet resulting in a model with $\cN=4$ deformed supersymmetry and
another $\cN=4$ hidden supersymmetry. However,  these supersymmetries
are global: they originate from the super-branes extended objects which
are known to interact with the supergravity satisfying classical field
equations. On shell background  supergravity is a condition for the
local fermionic $\kappa$-symmetry. Some new ideas will be required to
build the higher derivative superconformal invariants, if they exist, to
shed some light on the issue of the UV finiteness of extended
perturbative supergravity, as proposed in \cite{Ferrara:2012ui}.

\section{Acknowledgements}
We had stimulating discussions with S. Bellucci,  J. J. Carrasco, S.
Ferrara,  E. Ivanov, S. Kuzenko,   T. Ort{\'\i}n, M. Ro\v{c}ek, R. Roiban, A.
Tseytlin and Z. Komargodski. We are grateful to D. Sorokin and M. Tonin
for a discussion of the relation between  the hidden supersymmetry and
duality. This paper makes a small step towards clarifying this relation.

The work of R.K. was supported by SITP and NSF grant PHY-0756174 and by
the Templeton grant `Quantum Gravity Frontiers'. The work of E.B. and
C.S.S. and their visit to Stanford were partially supported by SITP and
by the Templeton grant `Quantum Gravity Frontiers'. The work of F.C. and
A.V.P. was supported in part by the FWO - Vlaanderen, Project No.
G.0651.11, and in part by the Interuniversity Attraction Poles Programme
initiated by the Belgian Science Policy (P7/37). E.B. and C.S.S. would
like to thank the Stanford Institute for Theoretical Physics, and F.C.
and A.V.P. the Centre for theoretical physics in Groningen for their
hospitality. E.B. and R.K. thank the Instituut voor Theoretische Fysica
in Leuven, where this work was completed, and in particular in
restaurant de Valck. This work has been supported in part by the Spanish
Ministry of Science and Education grant FPA2012-35043-C02-01, the
Comunidad de Madrid grant HEPHACOS S2009ESP-1473, and the Spanish
Consolider-Ingenio 2010 program CPAN CSD2007-00042. The work of C.S.S.
has been supported by the JAE-predoc grant JAEPre 2010 00613.


\appendix

\section{Appendix: Dp-branes}
\label{app:A}


Here we describe Dp superbranes in IIB theory with positive chiral
spinors.


\subsection{Dp-superbrane with local \texorpdfstring{$\kappa$}{kappa}-symmetry, maximal  supersymmetry and general covariance}
\label{ss:Dpmax}

The $\kappa$-symmetric Dp-brane action (with $p=2n+1$ odd), in a flat
background geometry with (longitudinal and transverse) coordinates
$X^{m}$, $m = 0,\dots , 9$, consists of the Dirac-Born-Infeld-Nambu-Goto
term $S_{\rm DBI}$ and Wess-Zumino term $S_{\rm WZ}$ in the world-volume
coordinates $\sigma^{\mu}$ $(\mu =0,\dots,p)$:
\begin{equation}
\label{actiongeneral}
S_{\rm DBI} +S_{\rm WZ} =  -\frac{1}{\alpha^2}
\int \rmd^{p+1} \sigma\, \sqrt{- \det (G_{\mu\nu} + \alpha
{\cal F}_{\mu\nu})} +\frac{1}{\alpha^2}\int \Omega_{p+1} \,.
\end{equation}
Here $G_{\mu\nu}$ is the  manifestly supersymmetric induced world-volume
metric\footnote{We use a doublet $\theta^{I},~I=1,2$ of Majorana-Weyl
spinors of the same chirality.  $\sigma _3$ in (\ref{defcalFbrane}) and
below indicates a Pauli $(\sigma _3)_I{}^J$ matrix (or any other
projection matrix). If it is clear by the context, we will omit the $I$
index as well as any spinorial index.}
\begin{equation}
G_{\mu\nu} = \eta_{mn} \Pi_\mu^m \Pi_\nu^n \ , \qquad \Pi_\mu^m =
\partial_\mu X^m - \bar\theta \Gamma^m \partial_\mu \theta \,,
\end{equation}
and the Born-Infeld field strength ${\cal F}_{\mu\nu}$ is
given by
\begin{equation}
{\cal F}_{\mu\nu} \equiv F_{\mu\nu} - b_{\mu\nu} \,, \qquad
b_{\mu\nu} = \alpha^{-1} \bar{\theta} \sigma_3
\Gamma_{m}\partial_{\mu}\theta\left(\partial_{\nu} X^{m}
-\frac{1}{2} \bar{\theta}\Gamma^{m}\partial_{\nu}\theta\right)-\left(
\mu\leftrightarrow \nu \right)\, ,
\label{defcalFbrane}
\end{equation}
where $\Omega_{p+1}$ is a particular $p+1$-form
\cite{Bergshoeff:1996tu,Aganagic:1996nn,Bergshoeff:1997kr} (see e.g.
(45) in \cite{Aganagic:1996nn}).

The action has the global supersymmetry
\begin{eqnarray}
\label{susytrans} \delta_\epsilon \theta &=& \epsilon\, , \qquad
\delta_\epsilon X^m = -\bar\theta\Gamma^m\epsilon\, , \qquad
\nonumber\\ [.2truecm]\label{a: superDbranes}
\delta_\epsilon A_\mu &=&-\alpha^{-1} \bar\theta \Gamma_m \sigma_3\epsilon
\partial_\mu X^m + \tfrac{\alpha^{-1}}{6} \left(\bar \theta\sigma_3 \Gamma_m \epsilon \bar\theta \Gamma^m
\partial_\mu\theta + \bar \theta\Gamma_m \epsilon\bar\theta \sigma_3 \Gamma^m \partial_\mu\theta\right)\,.
\end{eqnarray}
Note that as a consequence of the transformations above we have that
\begin{equation}
\delta_{\epsilon}\mathcal{F} = 0 \,.
\end{equation}
Also note that to show invariance of the action, we need the following
$d=10$ Fierz identity valid for any three Majorana-Weyl spinors
$\lambda_1,\lambda_2,\lambda_3$ of the same chirality
\begin{equation}
\Gamma_m\lambda_1\bar{\lambda}_2\Gamma^m\lambda_3+\Gamma_m\lambda_2\bar{\lambda}_3\Gamma^m\lambda_1+\Gamma_m\lambda_3\bar{\lambda}_1\Gamma^m\lambda_2=0\,.
\label{Fierzid10}
\end{equation}
Besides the global supersymmetry the action is also
invariant under a  local $\kappa$-symmetry given by
\begin{eqnarray}
 \delta_{\kappa} X^m &=&\bar\theta
\Gamma^m (\mathbb{1}+\Gamma)\kappa\,,
 \qquad \delta_{\kappa} \theta = (\mathbb{1}+\Gamma)\kappa \, ,\nonumber\\[.2truecm]
 \delta_{\kappa} A_\mu &=&  \alpha^{-1}
\bar\theta \sigma_3
\Gamma_m \delta_\kappa\theta \partial_\mu X^m \nonumber\\
&&- \tfrac{1}{2}\alpha^{-1}
\bar\theta \sigma_3 \Gamma_{m}
\delta_\kappa\theta \bar \theta \Gamma^m \partial_\mu\theta - \tfrac{1}{2}\alpha^{-1}
 \bar\theta \Gamma^m \delta_\kappa\theta \bar\theta \sigma_3 \Gamma_m \partial_\mu\theta\,,
\label{kappatrans}
\end{eqnarray}
where $\kappa(\sigma)\rightarrow\kappa^I(\sigma)$ is
an arbitrary doublet of Majorana-Weyl spinors of the same chirality.
$\Gamma$ is a hermitian traceless product structure, \emph{i.e.}, it
satisfies
\begin{equation}
\label{Gamma} \Tr\Gamma = 0\,, \qquad \Gamma^2 = 1\, .
\end{equation}
The precise form of $\Gamma$ can be found by explicitly
imposing $\kappa$-symmetry in (\ref{actiongeneral}). In the
usual Pauli matrices basis, and acting on positive-chirality spinors
$\theta^I$, $\Gamma$ is given by
\begin{equation}
\Gamma = \left(
\begin{array}{cc}
0 & \betaminus  \\
(-1)^n\betaplus & 0  \end{array} \right)\,. \label{defGamma}
\end{equation}
where $\betaplus$ and $\betaminus$ are matrices that satisfy
\begin{equation}
\betaminus\betaplus=\betaplus\betaminus=(-1)^n\,.
\label{Gbetabeta}
\end{equation}
They are constructed using pull-backs of the matrices $\Gamma ^m$ to the
world-volume:
\begin{equation}
\hat \Gamma_\mu \equiv \Pi_\mu{}^m \Gamma_m\,,\qquad  \hat{\Gamma }^\mu \equiv G^{\mu \nu
}\hat \Gamma_\nu = \Pi
^\mu _m\Gamma ^m\,,\qquad \Pi ^\mu _m= G^{\mu \nu }\Pi_\nu^n \eta_{mn}\,,
 \label{defhatGamma}
\end{equation}
where $G^{\mu \nu }$ is the inverse of $G_{\mu \nu }$. They satisfy the
Clifford algebra relations $\hat{\Gamma }^\mu\hat{\Gamma
}^\nu+\hat{\Gamma }^\nu\hat{\Gamma }^\mu=2G^{\mu \nu }$. We have
\begin{equation}
  \Pi^\mu_m \Pi^m_\nu =
\delta _\nu^\mu\,,
 \label{Piinversem}
\end{equation}
but since $\Pi $ is only a square matrix for $p=9$,  $\Pi ^\mu _m$ is
only the inverse matrix of $\Pi^m_\mu$ in that case.

In terms of the pull-backs, the matrices $\betaplus$ and $\betaminus$
are given by
\begin{eqnarray}
\betaplus &\equiv&\cG\,
se^{\frac{\alpha}{2}\mathcal{F}_{\mu\nu}\hat \Gamma^{\mu\nu}}\Gamma_{(0)}^{Dp}
\equiv \cG\,\sum^{n+1}_{k=0} \frac{\alpha^k}{2^k k!} \hat \Gamma^{\mu_1
\nu_1 \cdots \mu_k \nu_k }\mathcal{F}_{\mu_1 \nu_1}\cdots
\mathcal{F}_{\mu_k \nu_k}\Gamma_{(0)}^{Dp}\,,\nonumber\\
\betaminus &\equiv&\cG\,
se^{-\frac{\alpha}{2}\mathcal{F}_{\mu\nu}\hat \Gamma^{\mu\nu}}\Gamma_{(0)}^{Dp}
\equiv \cG\,\sum^{n+1}_{k=0} \frac{(-\alpha)^k}{2^k k!} \hat \Gamma^{\mu_1
\nu_1 \cdots \mu_k \nu_k }\mathcal{F}_{\mu_1 \nu_1}\cdots
\mathcal{F}_{\mu_k \nu_k}\Gamma_{(0)}^{Dp}\,.
\label{defbetaplusmin}
\end{eqnarray}
with
\begin{equation}
\cG=\frac{\sqrt{\left|G\right|}}{\sqrt{\left| G + \alpha \cal{F}\right|}}=
\left[ \det\left( \delta _\mu {}^\nu +\alpha {\cal F}_{\mu \rho }G^{\rho \nu }\right) \right] ^{-1/2}\,.
\label{defGF}
\end{equation}

Here $se$ stands for the skew-exponential function (i.e.~the exponential
function with skew-symmetrized indices of the gamma matrices at every
order in the expansion so the expansion has effectively only a finite
number of terms). $\beta _-$ is related to $\beta _+$ by replacing
${\cal F}$ by $-{\cal F}$.

The matrix $\Gamma_{(0)}^{Dp}$ is defined by
\begin{equation}
  \Gamma_{(0)}^{Dp} =
\frac{1}{(p+1)!\sqrt{|G|}}\varepsilon^{\mu_1\dots \mu_{p+1}}\hat \Gamma_{\mu_1\dots
\mu_{p+1}}\,,\qquad(\Gamma_{(0)}^{Dp})^2=(-1)^n\,.
\label{Gamma0Dp}
\end{equation}
In case of $p=9$ it agrees with the matrix defined in flat gamma
matrices:
\begin{equation}
 \Gamma_{(0)}^{D9}= \frac{1}{10!\sqrt{|G|}}\varepsilon^{\mu_0\dots \mu_{9}}\hat \Gamma_{\mu_0\dots
\mu_{9}}=\frac{1}{10!}\varepsilon^{m_0\dots m_{9}} \Gamma_{m_0\dots m_{9}}=\Gamma _*\,, \label{Gamma0D9}
\end{equation}

The theory is also invariant under general coordinate transformations
$\sigma^{\mu\, \prime} = \sigma^{\mu}+\xi^{\mu}(\sigma)$ on the
worldvolume. The complete set of transformations  on the fields of the
theory is hence given by supersymmetry transformations
$\delta_{\epsilon}$, $\kappa$-transformations $\delta_{\kappa}$,
covariant general coordinate transformations on the worldvolume
$\delta_{\xi}$, and a $U(1)$ gauge transformation $\delta_{\rm U(1)} $:
\begin{eqnarray}
\delta\theta &=& \delta_{\epsilon}\theta +
\delta_{\kappa} \theta + \xi^{\mu}\partial_{\mu} \theta\,, \nonumber \\
\delta X^m &=& \delta_{\epsilon}X^m + \delta_{\kappa}
X^m + \xi^{\mu}\partial_{\mu} X^m\,, \nonumber \\
\delta A_{\mu} &=&  \delta_{\epsilon} A_{\mu} +
\delta_{\kappa} A_{\mu} +{\delta}_{\rm U(1)} A_{\mu}+
\xi^{\rho} F_{\rho\mu}\ 
\,. \label{deltas}
\end{eqnarray}


\subsection{Gauge-fixing the Dp-superbrane}
\label{ss:gf}


Upon gauge fixing of $\kappa$-symmetry and general
coordinate transformations, the global target space supersymmetry
combines with a special field dependent $\kappa$-transformation into
a global worldvolume supersymmetry. Writing
\begin{equation}
X^m =\{X^{m^\prime}, \phi^I\}\,, \qquad  m^\prime =0,1,\dots,p\,,\quad  I=1,\dots,9-p\,,
\label{Xmtoprimeandphi}
\end{equation}
where $m^\prime$ refers to the $p+1$
worldvolume directions and $I$ refers to the $9-p$ transverse
directions, we impose the following gauge-fixing conditions
\begin{equation}
\left(\mathbb{1}+\sigma_3\right)\theta = 0\, ,
\qquad X^{m^\prime} =\delta^{m^\prime}_{\mu}\sigma^{\mu} \,.
\label{gaugefixing3}
\end{equation}
Using the basis where $\sigma_3$ is the
diagonal Pauli matrix, the condition (\ref{gaugefixing3}) implies
$\theta^{1} = 0$. From now on we will use  $\theta^2 \equiv
\alpha\lambda$.

In this gauge the Wess-Zumino term $\Omega _{p+1}$ becomes constant
($-1$) and the action is given by
\begin{equation}
S =  -\frac{1}{\alpha^2} \int \rmd^{p+1} \sigma\,\left\{
\sqrt{- \det (G_{\mu\nu} + \alpha {\cal F}_{\mu\nu})} - 1 \right\}\,,
\label{actionBI}
\end{equation}
together with
\begin{eqnarray}
G_{\mu\nu} &=& \eta_{m^\prime n^\prime} \Pi_\mu^{m^\prime}
\Pi_\nu^{n^\prime} +\delta_{IJ} \Pi_\mu^I\Pi_\nu^J \ ,\nonumber \\
[.2truecm] \Pi_\mu^{m^\prime} &=& \delta^{m^\prime}_{\mu}
-\alpha^2\bar\lambda \Gamma^{m^\prime}
\partial_\mu \lambda \ , \qquad \Pi_\mu^I = \partial_\mu\phi^I
-\alpha^2\bar\lambda \Gamma^I
\partial_\mu \lambda\,,
\end{eqnarray}
and
\begin{eqnarray}
{\cal F}_{\mu\nu} &\equiv& F_{\mu\nu} - b_{\mu\nu} \ ,
\nonumber\\ [.2truecm] b_{\mu\nu} &=&
-2\alpha\bar{\lambda}\Gamma_{[\nu }\partial_{\mu]}\lambda\,
-2\alpha\bar{\lambda}\Gamma_{I}\partial_{[\mu}\lambda\,\partial_{\nu]}
\phi^I\, .
\end{eqnarray}
In order to preserve the gauge-fixing conditions
(\ref{gaugefixing3}) we have to impose
\begin{equation}
\delta \theta^{1}=0\, , \qquad \delta X^{m^\prime} = 0\,,
\end{equation}
from which we obtain
\begin{equation}
\epsilon^1+\kappa^1+\betaminus\kappa^2=0\, \qquad \text{and} \qquad
\xi^{m^\prime} =\alpha \left\{
\bar{\lambda}\Gamma^{m^\prime}\epsilon^2 +(-1)^n
\bar{\lambda}\Gamma^{m^\prime}\betaplus
\epsilon^{1} \right\}\,, \label{decompositionxi}
\end{equation}
together with the corresponding gauge-fixed, non-linear
realization of the remaining transformations
\begin{eqnarray}
\delta_{\epsilon^{1,2}} \phi^I &=& -\alpha\bar\lambda\Gamma^I{\epsilon}^2 -(-)^n\alpha
\bar\lambda\Gamma^I\betaplus\epsilon^1
+\xi^\mu\partial_\mu \phi^I\,, \nonumber \\
\delta_{\epsilon^{1,2}} \lambda &=& \alpha^{-1}\left(\epsilon^2 +(-1)^{n+1}
\betaplus\epsilon^{1}\right)
+\xi^{\mu}\partial_{\mu}\lambda\,, \nonumber \\
 \delta_{\epsilon^{1,2}} A_\mu&=&\bar \lambda\big(
 \Gamma_\mu + \Gamma_I\partial_\mu\phi^I\big) \left(\epsilon^2 +(-)^n \betaplus\epsilon^1 \right)
 \nonumber \\
&&- \alpha^2 \bar \lambda \Gamma_m(\tfrac{1}{3} \epsilon^2 + (-)^n\betaplus\epsilon^1
) \bar\lambda \Gamma^m \partial_\mu \lambda+\xi^\rho F_{\rho\mu} \,. \label{decompositionsgf}
\end{eqnarray}
The above transformations are parametrized by two spinors
$\epsilon^{1}$ and $\epsilon^{2}$, and therefore we have two
independent sets of sixteen supersymmetry transformations. The deformed linear
symmetries and the VA-type
non-linear ones are given by the change of basis
\begin{equation}
  \epsilon^1=-\tfrac{1}{2}\rmi^n\Gamma _*^{(p)}\epsilon\,,\qquad\epsilon^2=-\tfrac{1}{2}\epsilon+\zeta\,.
 \label{redefeps12epszeta}
\end{equation}
where $\Gamma _*^{(p)}=(-\rmi)^{n}\Gamma ^0\ldots \Gamma ^{p}$ (and in
particular $\Gamma _*^{(9)}=\Gamma _*$). Using this in
(\ref{decompositionxi}) gives
\begin{equation}
  \xi ^{m}=\xi _\epsilon ^{m}+\xi _\zeta ^{m}\,,\qquad\xi _\epsilon ^{m} \equiv
-\tfrac{1}{2}\alpha\bar{\lambda}\Gamma^{m}\left[\mathbb{1} +
\betaplusprime\right]\epsilon\,,\qquad\xi _\zeta ^{m}=\alpha \bar\lambda \Gamma ^{m} \zeta\,,
 \label{xiepszeta}
\end{equation}
where
\begin{equation}
  \betaplusprime=(-\rmi)^n\betaplus\Gamma _*^{(p)}=(-\rmi)^n\cG\sum^{n+1}_{k=0} \frac{\alpha^k}{2^k k!} \hat \Gamma^{\mu_1
\nu_1 \cdots \mu_k \nu_k }\mathcal{F}_{\mu_1 \nu_1}\cdots
\mathcal{F}_{\mu_k \nu_k}\Gamma _{(0)}^{Dp}\Gamma _*^{(p)}=1+{\cal O}(\alpha )\,,
 \label{betaplusprime}
\end{equation}
when terms with $\partial  _\mu \phi ^I$ are also considered as order
$\alpha $ (see e.g. (\ref{philambda10to4})). The corresponding
transformations are
\begin{eqnarray}
\delta_\epsilon  \phi^I &=& \ft12\alpha\bar\lambda\Gamma^I\left[\unity +
\betaplusprime\right] \epsilon
+\xi^\mu_\epsilon \partial_\mu \phi^I\,, \nonumber \\
\delta_\epsilon  \lambda &=&- \ft1{2\alpha}\left[\unity -
\betaplusprime\right]\epsilon
+\xi^{\mu}_\epsilon \partial_{\mu}\lambda\,, \nonumber \\
 \delta_\epsilon  A_\mu&=&-\ft12\bar \lambda\big(
 \Gamma_\mu + \Gamma_I\partial_\mu\phi^I\big) \left[\unity +\betaplusprime\right]\epsilon
 \nonumber \\
&&+\ft12 \alpha^2 \bar \lambda \Gamma_m\left[ \tfrac{1}{3}\unity  + \betaplusprime\right] \epsilon
 \bar\lambda \Gamma^m \partial_\mu \lambda+\xi_\epsilon ^\rho F_{\rho\mu} \,. \label{decompositionsgfeps}
\end{eqnarray}
and
\begin{eqnarray}
\delta_\zeta   \phi^I &=& -\alpha\bar\lambda\Gamma^I\zeta
+\xi^\mu_\zeta \partial_\mu \phi^I\,, \nonumber \\
\delta_\zeta   \lambda &=& \alpha^{-1}\zeta
+\xi^{\mu}_\zeta \partial_{\mu}\lambda\,, \nonumber \\
 \delta_\zeta   A_\mu&=&\bar \lambda\big(
 \Gamma_\mu + \Gamma_I\partial_\mu\phi^I\big) \zeta
- \tfrac{1}{3}\alpha^2 \bar \lambda \Gamma_m \zeta
 \bar\lambda \Gamma^m \partial_\mu \lambda+\xi^\rho_\zeta  F_{\rho\mu} \,. \label{decompositionsgfzeta}
\end{eqnarray}

\section{Appendix: Vp-branes}
\label{app:B}

\subsection{Vector p-branes in \texorpdfstring{$d=6$}{d=6}}

In the previous appendix we constructed  supersymmetric Born-Infeld
actions with 16+16 supersymmetries corresponding to Dp-branes in $d=10$.
It is the purpose of this section to give similar results for Vp-branes
with 8+8 supersymmetries in $d=6$. We remind that Vp-branes are branes
whose worldvolume content is given by a vector multiplet but whose
tension is not necessarily proportional to the inverse string coupling
constant. Since the discussion in this appendix is rather similar to the
previous appendix we will only highlight a few relevant formulae.

The $\kappa$-symmetric Vp-brane action (with $p=2n+1$ odd), in a flat
background geometry with (longitudinal and transverse) coordinates
$X^{m}$, $m = 0,\dots , 5$, consists of the Dirac-Born-Infeld-Nambu-Goto
term $S_{\text{DBI}}$ and Wess-Zumino term $S_{\rm WZ}$ and is formally
completely the same as (\ref{actiongeneral}). All formulas of Sec.
\ref{ss:Dpmax} are still applicable.\footnote{In this case, we use a
doublet $\theta^{I i},~I=1,2$ of symplectic Majorana-Weyl spinors of the
same chirality. The symplectic indices $i$ are implicitly present in the
same way as the index $I$. In a bilinear we assume that they are
contracted using the antisymmetric $\varepsilon_{ij}$, e.g.
$\bar{\lambda}\Gamma^m\chi\equiv\bar{\lambda}^i\Gamma^m\chi^j\varepsilon_{ji}$.
}

To show invariance of the action, we need in this case the $d=6$ Fierz
identity valid for any three symplectic Majorana-Weyl spinors
$\lambda_1^i,\lambda_2^j,\lambda_3^k$ of the same chirality, that is
also formally the same as  (\ref{Fierzid10}), or explicitly
\begin{equation}
\Gamma_m\lambda_1^i\bar{\lambda}_2^j\Gamma^m\lambda_3^k\varepsilon _{kj}
+\Gamma_m\lambda_2^i\bar{\lambda}_3^j\Gamma^m\lambda_1^k\varepsilon _{kj}
+\Gamma_m\lambda_3^i\bar{\lambda}_1^j\Gamma^m\lambda_2^k\varepsilon _{kj}=0\,.
\label{Fierzcyclicd6}
\end{equation}

Also the global $\epsilon $ and local $\kappa $-supersymmetry have the
same form as in (\ref{susytrans}) and  (\ref{kappatrans}), where
$\kappa(\sigma)\rightarrow\kappa^{Ii}(\sigma)$ is an arbitrary doublet
of symplectic Majorana-Weyl spinors of the same chirality, and
the $\Gamma^m$ and $\Pi_{\mu}^m$ are, of course, the $d=6$ quantities.

The theory is also invariant under general coordinate
transformations $\sigma^{\mu\, \prime} =
\sigma^{\mu}+\xi^{\mu}(\sigma)$ on the worldvolume. Hence, just as in (\ref{deltas}) of the previous appendix, the
complete set of transformations  on the fields of the theory is
given by supersymmetry transformations,
$\kappa$-transformations, general coordinate
transformations on the worldvolume and a $U(1)$
gauge transformation.

\subsection{Gauge fixing}

The gauge fixing of $\kappa$-symmetry and general coordinate
transformations goes exactly as in Sec. \ref{ss:gf}, and the formulas
remain applicable, with the understanding that now $I=1,\dots,5-p$.

The decomposition rules and non-linear gauge-fixed realization of the
remaining transformations is analogous as in (\ref{decompositionxi}) and
(\ref{decompositionsgf}). Also the split into two independent sets of,
in this case eight, supersymmetry transformations, is exactly as in
(\ref{decompositionsgfeps}) and (\ref{decompositionsgfzeta}).

\section{Notation}
\label{app:notation}

We follow the notation of \cite{Freedman:2012zz}, especially all
coefficients in Appendix A in that book are $s_i=1$ except for the
normalization of the $\epsilon $-supersymmetry, which is such that
$s_9=-2$. One can go to the standard notations of the book by
multiplying all barred spinors by $-1/2$, e.g. the standard kinetic
Lagrangian for the real fermion becomes $-\ft12\bar \lambda
\slashed{\partial }\lambda $ rather than $\bar \lambda \slashed{\partial
}\lambda $. The $\gamma $-matrices for $d=10$ or $d=6$ are denoted as
$\Gamma ^m$. Note that the matrices $\Gamma^m$ and $\Gamma_\mu$ are the
flat $d=10$ matrices, whereas $ \hat \Gamma_{\mu}$ is defined in
(\ref{defhatGamma}) as the pull-back to the word-volume:
\begin{equation}
\Gamma_{\mu} = \delta^m_{\mu}\Gamma_{m}\,,\qquad
\hat \Gamma_{\mu} = \Pi^m_{\mu}\Gamma_{m}= (\delta_\mu^m - \alpha^2\bar\lambda \Gamma^m \partial_\mu \lambda) \Gamma_m\,.
\label{gamma}
\end{equation}

We use shortcuts for index contractions, such that $F^2= F_{\mu \nu
}F^{\mu \nu }$ and $\Gamma \cdot F= \Gamma^{\mu \nu }F_{\mu \nu }$.
However, matrices $F^3$, \ldots  are defined as $(F^3)_{\mu\nu} =F_{\mu
\rho }F^{\rho \sigma }F_{\sigma \nu }$, and $\Tr F^4= F_{\mu \nu }F^{\nu
\rho }F_{\rho \sigma }F^{\sigma \mu }$.

As in \cite{Freedman:2012zz}, we define
\begin{equation}
  \Gamma _*=(-\rmi)^{(d-2)/2}\Gamma ^0\ldots \Gamma ^{d-1}\,.
 \label{gammastar}
\end{equation}
The spinors for $d=10$ and $d=6$ are left-handed, which means that
\begin{eqnarray}
  &&\Gamma _*\lambda =\lambda \,,\qquad \bar \lambda \Gamma _*=-\bar \lambda\,, \label{deflefthanded}\\
 &&d=10:\ \Gamma _*=\Gamma ^0\Gamma ^1\ldots \Gamma ^9\,,\qquad
 d=6:\  \Gamma _*=-\Gamma ^0\Gamma ^1\ldots \Gamma ^5 
 \,.\nonumber
\end{eqnarray}

In this work appear Majorana and symplectic-Majorana spinors that
satisfy the Majorana flip relations
\begin{eqnarray}
  &&\bar \lambda _1 \lambda _2= \bar \lambda _2 \lambda _1\,,\qquad
  \bar \lambda _1 \Gamma ^\mu \lambda _2=- \bar \lambda _2 \Gamma ^\mu  \lambda _1\,,\qquad
\bar \lambda _1 \Gamma ^{\mu\nu } \lambda _2=- \bar \lambda _2 \Gamma ^{\mu\nu }  \lambda _1\,,\qquad
\bar \lambda _1 \Gamma ^{\mu\nu\rho  } \lambda _2=\bar \lambda _2 \Gamma ^{\mu\nu\rho  }  \lambda _1\,,\nonumber\\
&&\bar \lambda _1 \Gamma ^{\mu_1\ldots \mu _r  } \lambda _2=(-)^{r(r+1)/2}\bar \lambda _2 \Gamma ^{\mu_1\ldots \mu _r   }  \lambda _1
 \label{flip}
\end{eqnarray}
For $d=6$ this is accomplished due to a definition where the indices
$i,j,\ldots =1,2$ are hidden and
\begin{equation}
\bar{\lambda}\Gamma^\mu \chi\equiv\bar{\lambda}^i\Gamma^\mu \chi^j\epsilon_{ji}\,.
 \label{symplecticcontraction}
\end{equation}
In this way the formulas can be used for all dimensions $d=2,3,4,6$ and
10, where also the cyclic identity (\ref{Fierzid10}) holds, which for
$d=6$ is given explicitly in (\ref{Fierzcyclicd6}).

To reduce the $d=10$ expressions to $d=4$ spinors in Sec.
\ref{ss:Maxwd4N4}, we construct the  $32\times 32$  matrices $\Gamma^m
$, with $m=\mu ,a,a+3$ ($\mu =0,\ldots 3$ and $a=1,2,3$) from $4\times
4$ matrices $\gamma^\mu $ by
\begin{eqnarray}
 \Gamma ^\mu  & = & \gamma ^\mu \otimes \unity _8 \,,\qquad
 \Gamma ^a = \gamma _*\otimes \begin{pmatrix}0&\beta ^a \cr -\beta ^a&0\end{pmatrix}\,,\qquad
 \Gamma ^{a+3}=\gamma _*\otimes \begin{pmatrix}0&\rmi\alpha  ^a \cr \rmi\alpha  ^a&0\end{pmatrix}\,,\nonumber\\
 C_{10}&=&C_4\otimes\begin{pmatrix}0&\unity _4 \cr \unity _4&0\end{pmatrix}\,,\qquad
 \Gamma _*=\gamma _*\otimes \begin{pmatrix}\unity _4&0 \cr 0&-\unity _4\end{pmatrix}\,,
 \label{Gamma104}
\end{eqnarray}
where $C_{10}$ is the charge conjugate in 10 dimensions, and $C_4$ the
one of $d=4$. Here we use the $4\times 4$ antisymmetric real matrices
$\alpha ^a$ and $\beta ^a$ from \cite{Brink:1976bc,Gliozzi:1976qd},
whose components are indicated by indices $i,j=1,\ldots ,4$, and which
satisfy
\begin{eqnarray}
{} \{\alpha ^a,\alpha ^b\} & = & \{\beta ^a,\beta  ^b\}=-2\delta ^{ab}\,,\qquad
{}[\alpha ^a,\beta  ^b]=0\,,\nonumber\\
\alpha ^a_{ij} \alpha ^a_{k\ell }&=& 2\delta _{i[k}\delta _{\ell]j }+\varepsilon _{ijk\ell }\,,\qquad
\beta  ^a_{ij} \beta ^a_{k\ell }= 2\delta _{i[k}\delta _{\ell]j }-\varepsilon _{ijk\ell }\,,\nonumber\\
\alpha ^a_{ij}&=&\ft12\varepsilon _{ijk\ell }\alpha ^a_{k\ell }\,,\qquad\beta  ^a_{ij}=-\ft12\varepsilon _{ijk\ell }\beta  ^a_{k\ell }\,,\qquad
\alpha ^a_{ij}\alpha ^{b\,ij}=\beta  ^a_{ij}\beta  ^{b\,ij}=4\delta ^{ab}\,,\nonumber\\
\alpha ^1\alpha ^2\alpha ^3&=&\unity _4\,,\qquad \beta ^1\beta ^2\beta ^3=-\unity _4\,.
 \label{propalbe}
\end{eqnarray}

The engineering dimensions that are used for the various fields in a
$d$-dimensional action are
\begin{eqnarray}
&&[x]=-1\,,\qquad  [\theta, \epsilon, \eta,\zeta,\kappa  ]=-1/2\,, \qquad [A_\mu]=d/2-1\,, \qquad [F_{\mu\nu}]=d/2\,,\qquad  [\lambda
]=(d-1)/2\,,\nonumber\\
&& [\alpha] = -d/2\,,\qquad [\partial _\mu ]=1\,,\qquad [X]=-1\,,\qquad [b_{\mu \nu }]=d/2\,,\qquad [\phi ^I]=-1\,.
 \label{dimensionsd}
\end{eqnarray}


\begin{thebibliography}{10}

\bibitem{Ferrara:2012ui}
S.~Ferrara, R.~Kallosh  and A.~Van~Proeyen, \emph{{Conjecture on hidden
  superconformal symmetry of $N=4$ Supergravity}}, Phys.Rev. {\bf D87} (2013)
  \href{http://dx.doi.org/10.1103/PhysRevD.87.025004}{025004},
\href{http://arxiv.org/abs/1209.0418}{{\tt arXiv:1209.0418 [hep-th]}}

\bibitem{deWit:2010za}
B.~de~Wit, S.~Katmadas  and M.~van Zalk, \emph{{New supersymmetric
  higher-derivative couplings: Full $N=2$ superspace does not count!}}, JHEP
  {\bf 1101} (2011) \href{http://dx.doi.org/10.1007/JHEP01(2011)007}{007},
\href{http://arxiv.org/abs/1010.2150}{{\tt arXiv:1010.2150 [hep-th]}}

\bibitem{Chemissany:2012pf}
W.~Chemissany, S.~Ferrara, R.~Kallosh  and C.~Shahbazi, \emph{{$N=2$
  supergravity counterterms, off and on shell}}, JHEP {\bf 12} (2012)
  \href{http://dx.doi.org/10.1007/JHEP12(2012)089}{089},
\href{http://arxiv.org/abs/1208.4801}{{\tt arXiv:1208.4801 [hep-th]}}

\bibitem{Bergshoeff:1986jm}
E.~Bergshoeff, M.~Rakowski  and E.~Sezgin, \emph{{Higher derivative
  super-Yang-Mills theories}}, Phys.Lett. {\bf B185} (1987)
\href{http://dx.doi.org/10.1016/0370-2693(87)91017-3}{371}

\bibitem{Born:1934gh}
M.~Born and L.~Infeld, \emph{{Foundations of the new field theory}},
  Proc.Roy.Soc.Lond. {\bf A144} (1934)
425--451

\bibitem{Dirac:1962iy}
P.~A. Dirac, \emph{{An Extensible model of the electron}},
Proc.Roy.Soc.Lond.
  {\bf A268} (1962)
57--67

\bibitem{Deser:1980ck}
S.~Deser and R.~Puzalowski, \emph{{Supersymmetric nonpolynomial vector
  multiplets and causal propagation}}, J.Phys. {\bf A13} (1980)
\href{http://dx.doi.org/10.1088/0305-4470/13/7/031}{2501}

\bibitem{Cecotti:1986gb}
S.~Cecotti and S.~Ferrara, \emph{{Supersymmetric Born-Infeld
Lagrangians}},
  Phys.Lett. {\bf B187} (1987)
\href{http://dx.doi.org/10.1016/0370-2693(87)91105-1}{335}

\bibitem{Metsaev:1987qp}
R.~Metsaev, M.~Rakhmanov  and A.~A. Tseytlin, \emph{{The Born-Infeld
action as
  the effective action in the open superstring theory}}, Phys.Lett. {\bf B193}
  (1987)
\href{http://dx.doi.org/10.1016/0370-2693(87)91223-8}{207}

\bibitem{Bagger:1996wp}
J.~Bagger and A.~Galperin, \emph{{A new Goldstone multiplet for
partially
  broken supersymmetry}}, Phys.Rev. {\bf D55} (1997)
  \href{http://dx.doi.org/10.1103/PhysRevD.55.1091}{1091--1098},
\href{http://arxiv.org/abs/hep-th/9608177}{{\tt arXiv:hep-th/9608177
[hep-th]}}

\bibitem{Rocek:1997hi}
M.~Ro\v{c}ek and A.~A. Tseytlin, \emph{{Partial breaking of global $D =
4$
  supersymmetry, constrained superfields, and three-brane actions}}, Phys.Rev.
  {\bf D59} (1999) \href{http://dx.doi.org/10.1103/PhysRevD.59.106001}{106001},
\href{http://arxiv.org/abs/hep-th/9811232}{{\tt arXiv:hep-th/9811232
[hep-th]}}

\bibitem{Tseytlin:1999dj}
A.~A. Tseytlin, \emph{{Born-Infeld action, supersymmetry and string
theory}},
  \href{http://arxiv.org/abs/hep-th/9908105}{{\tt arXiv:hep-th/9908105
  [hep-th]}},
In Shifman, M.A. (ed.): The many faces of the superworld, 417-452

\bibitem{Ketov:1998ku}
S.~V. Ketov, \emph{{A manifestly $N=2$ supersymmetric Born-Infeld
action}},
  Mod.Phys.Lett. {\bf A14} (1999)
  \href{http://dx.doi.org/10.1142/S0217732399000559}{501--510},
\href{http://arxiv.org/abs/hep-th/9809121}{{\tt arXiv:hep-th/9809121
[hep-th]}}

\bibitem{Bellucci:2000ft}
S.~Bellucci, E.~Ivanov  and S.~Krivonos, \emph{{$N=2$ and $N=4$
supersymmetric
  Born-Infeld theories from nonlinear realizations}}, Phys.Lett. {\bf B502}
  (2001) \href{http://dx.doi.org/10.1016/S0370-2693(01)00142-3}{279--290},
\href{http://arxiv.org/abs/hep-th/0012236}{{\tt arXiv:hep-th/0012236
[hep-th]}}

\bibitem{Ketov:2001dq}
S.~V. Ketov, \emph{{Many faces of Born-Infeld theory}},
\href{http://arxiv.org/abs/hep-th/0108189}{{\tt arXiv:hep-th/0108189
[hep-th]}}

\bibitem{Kuzenko:2000tg}
S.~M. Kuzenko and S.~Theisen, \emph{{Supersymmetric duality rotations}},
JHEP
  {\bf 0003} (2000) 034,
\href{http://arxiv.org/abs/hep-th/0001068}{{\tt arXiv:hep-th/0001068
[hep-th]}}

\bibitem{Kuzenko:2000uh}
S.~M. Kuzenko and S.~Theisen, \emph{{Nonlinear selfduality and
supersymmetry}},
  Fortsch.Phys. {\bf 49} (2001) 273--309,
\href{http://arxiv.org/abs/hep-th/0007231}{{\tt arXiv:hep-th/0007231
[hep-th]}}

\bibitem{Bellucci:2001hd}
S.~Bellucci, E.~Ivanov  and S.~Krivonos, \emph{{Towards the complete
$N=2$
  superfield Born-Infeld action with partially broken $N=4$ supersymmetry}},
  Phys.Rev. {\bf D64} (2001)
  \href{http://dx.doi.org/10.1103/PhysRevD.64.025014}{025014},
\href{http://arxiv.org/abs/hep-th/0101195}{{\tt arXiv:hep-th/0101195
[hep-th]}}

\bibitem{Kerstan:2002au}
S.~F. Kerstan, \emph{{Supersymmetric Born-Infeld from the D9-brane}},
  Class.Quant.Grav. {\bf 19} (2002)
  \href{http://dx.doi.org/10.1088/0264-9381/19/17/304}{4525--4536},
\href{http://arxiv.org/abs/hep-th/0204225}{{\tt arXiv:hep-th/0204225
[hep-th]}}

\bibitem{Ivanov:2002ab}
E.~Ivanov and B.~Zupnik, \emph{{New representation for Lagrangians of
selfdual
  nonlinear electrodynamics}},
\href{http://arxiv.org/abs/hep-th/0202203}{{\tt arXiv:hep-th/0202203
[hep-th]}}

\bibitem{Berkovits:2002ag}
N.~Berkovits and V.~Pershin, \emph{{Supersymmetric Born-Infeld from the
pure
  spinor formalism of the open superstring}}, JHEP {\bf 0301} (2003) 023,
\href{http://arxiv.org/abs/hep-th/0205154}{{\tt arXiv:hep-th/0205154
[hep-th]}}

\bibitem{Ivanov:2003uj}
E.~Ivanov and B.~Zupnik, \emph{{New approach to nonlinear
electrodynamics:
  Dualities as symmetries of interaction}}, Phys.Atom.Nucl. {\bf 67} (2004)
  \href{http://dx.doi.org/10.1134/1.1842299}{2188--2199},
\href{http://arxiv.org/abs/hep-th/0303192}{{\tt arXiv:hep-th/0303192
[hep-th]}}

\bibitem{Sorokin:1999jx}
D.~P. Sorokin, \emph{{Superbranes and superembeddings}}, Phys.Rept. {\bf
329}
  (2000) \href{http://dx.doi.org/10.1016/S0370-1573(99)00104-0}{1--101},
\href{http://arxiv.org/abs/hep-th/9906142}{{\tt arXiv:hep-th/9906142
[hep-th]}}

\bibitem{Pasti:2000zs}
P.~Pasti, D.~P. Sorokin  and M.~Tonin, \emph{{Superembeddings, partial
  supersymmetry breaking and superbranes}}, Nucl.Phys. {\bf B591} (2000)
  \href{http://dx.doi.org/10.1016/S0550-3213(00)00569-1}{109--138},
\href{http://arxiv.org/abs/hep-th/0007048}{{\tt arXiv:hep-th/0007048
[hep-th]}}

\bibitem{Carrasco:2011jv}
J.~J.~M. Carrasco, R.~Kallosh  and R.~Roiban, \emph{{Covariant
procedures for
  perturbative non-linear deformation of duality-invariant theories}},
  Phys.Rev. {\bf D85} (2012)
  \href{http://dx.doi.org/10.1103/PhysRevD.85.025007}{025007},
\href{http://arxiv.org/abs/1108.4390}{{\tt arXiv:1108.4390 [hep-th]}}

\bibitem{Chemissany:2011yv}
W.~Chemissany, R.~Kallosh  and T.~Ort{\'\i}n, \emph{{Born-Infeld with higher
  derivatives}}, Phys.Rev. {\bf D85} (2012)
  \href{http://dx.doi.org/10.1103/PhysRevD.85.046002}{046002},
\href{http://arxiv.org/abs/1112.0332}{{\tt arXiv:1112.0332 [hep-th]}}

\bibitem{Broedel:2012gf}
J.~Broedel, J.~J.~M. Carrasco, S.~Ferrara, R.~Kallosh  and R.~Roiban,
  \emph{{$N=2$ supersymmetry and U(1)-duality}}, Phys.Rev. {\bf D85} (2012)
  \href{http://dx.doi.org/10.1103/PhysRevD.85.125036}{125036},
\href{http://arxiv.org/abs/1202.0014}{{\tt arXiv:1202.0014 [hep-th]}}

\bibitem{Pasti:2012wv}
P.~Pasti, D.~Sorokin  and M.~Tonin, \emph{{Covariant actions for models
with
  non-linear twisted self-duality}}, Phys.Rev. {\bf D86} (2012)
  \href{http://dx.doi.org/10.1103/PhysRevD.86.045013}{045013},
\href{http://arxiv.org/abs/1205.4243}{{\tt arXiv:1205.4243 [hep-th]}}

\bibitem{Ivanov:2012bq}
E.~Ivanov and B.~Zupnik, \emph{{Bispinor auxiliary fields in
duality-invariant
  electrodynamics revisited}},
\href{http://arxiv.org/abs/1212.6637}{{\tt arXiv:1212.6637 [hep-th]}}

\bibitem{Kuzenko:2013gr}
S.~M. Kuzenko, \emph{{Duality rotations in supersymmetric nonlinear
  electrodynamics revisited}},
\href{http://arxiv.org/abs/1301.5194}{{\tt arXiv:1301.5194 [hep-th]}}

\bibitem{Aschieri:2013nda}
P.~Aschieri and S.~Ferrara, \emph{{Constitutive relations and
Schr{\"o}dinger's
  formulation of nonlinear electromagnetic theories}},
\href{http://arxiv.org/abs/1302.4737}{{\tt arXiv:1302.4737 [hep-th]}}

\bibitem{Volkov:1973ix}
D.~Volkov and V.~Akulov, \emph{{Is the neutrino a Goldstone particle?}},
Phys.
  Lett. {\bf B46} (1973)
\href{http://dx.doi.org/10.1016/0370-2693(73)90490-5}{109--110}

\bibitem{Cederwall:1996pv}
M.~Cederwall, A.~von Gussich, B.~E. Nilsson  and A.~Westerberg,
\emph{{The
  Dirichlet super three-brane in ten-dimensional type IIB supergravity}},
  Nucl.Phys. {\bf B490} (1997)
  \href{http://dx.doi.org/10.1016/S0550-3213(97)00071-0}{163--178},
\href{http://arxiv.org/abs/hep-th/9610148}{{\tt arXiv:hep-th/9610148
[hep-th]}}

\bibitem{Cederwall:1996ri}
M.~Cederwall, A.~von Gussich, B.~E. Nilsson, P.~Sundell  and
A.~Westerberg,
  \emph{{The Dirichlet super p-branes in ten-dimensional type IIA and IIB
  supergravity}}, Nucl.Phys. {\bf B490} (1997)
  \href{http://dx.doi.org/10.1016/S0550-3213(97)00075-8}{179--201},
\href{http://arxiv.org/abs/hep-th/9611159}{{\tt arXiv:hep-th/9611159
[hep-th]}}

\bibitem{Bergshoeff:1996tu}
E.~Bergshoeff and P.~Townsend, \emph{{Super D-branes}}, Nucl.Phys. {\bf
B490}
  (1997) \href{http://dx.doi.org/10.1016/S0550-3213(97)00072-2}{145--162},
\href{http://arxiv.org/abs/hep-th/9611173}{{\tt arXiv:hep-th/9611173
[hep-th]}}

\bibitem{Aganagic:1996nn}
M.~Aganagic, C.~Popescu  and J.~H. Schwarz, \emph{{Gauge invariant and
gauge
  fixed D-brane actions}}, Nucl.Phys. {\bf B495} (1997)
  \href{http://dx.doi.org/10.1016/S0550-3213(97)00180-6}{99--126},
\href{http://arxiv.org/abs/hep-th/9612080}{{\tt arXiv:hep-th/9612080
[hep-th]}}

\bibitem{Bergshoeff:1997kr}
E.~Bergshoeff, R.~Kallosh, T.~Ort{\'\i}n  and G.~Papadopoulos,
  \emph{{$\kappa$-symmetry, supersymmetry and intersecting branes}}, Nucl.Phys.
  {\bf B502} (1997)
  \href{http://dx.doi.org/10.1016/S0550-3213(97)00470-7}{149--169},
\href{http://arxiv.org/abs/hep-th/9705040}{{\tt arXiv:hep-th/9705040
[hep-th]}}

\bibitem{Kallosh:1997aw}
R.~Kallosh, \emph{{Volkov-Akulov theory and D-branes}},
\href{http://arxiv.org/abs/hep-th/9705118}{{\tt arXiv:hep-th/9705118
[hep-th]}}

\bibitem{Bergshoeff:2012jb}
E.~A. Bergshoeff and F.~Riccioni, \emph{{Heterotic wrapping rules}},
JHEP {\bf
  1301} (2013) \href{http://dx.doi.org/10.1007/JHEP01(2013)005}{005},
\href{http://arxiv.org/abs/1210.1422}{{\tt arXiv:1210.1422 [hep-th]}}

\bibitem{Casalbuoni:1988xh}
R.~Casalbuoni, S.~De~Curtis, D.~Dominici, F.~Feruglio  and R.~Gatto,
  \emph{{Non-linear realization of supersymmetry algebra from supersymmetric
  constraint}}, Phys.Lett. {\bf B220} (1989)
\href{http://dx.doi.org/10.1016/0370-2693(89)90788-0}{569}

\bibitem{Komargodski:2009rz}
Z.~Komargodski and N.~Seiberg, \emph{{From linear SUSY to constrained
  superfields}}, JHEP {\bf 0909} (2009)
  \href{http://dx.doi.org/10.1088/1126-6708/2009/09/066}{066},
\href{http://arxiv.org/abs/0907.2441}{{\tt arXiv:0907.2441 [hep-th]}}

\bibitem{Kuzenko:2011tj}
S.~M. Kuzenko and S.~J. Tyler, \emph{{On the Goldstino actions and their
  symmetries}}, JHEP {\bf 1105} (2011)
  \href{http://dx.doi.org/10.1007/JHEP05(2011)055}{055},
\href{http://arxiv.org/abs/1102.3043}{{\tt arXiv:1102.3043 [hep-th]}}

\bibitem{Beisert:2010jx}
N.~Beisert, H.~Elvang, D.~Z. Freedman, M.~Kiermaier, A.~Morales  and
  S.~Stieberger, \emph{{$E_{7(7)}$ constraints on counterterms in ${\cal N}=8$
  supergravity}}, Phys.Lett. {\bf B694} (2010)
  \href{http://dx.doi.org/10.1016/j.physletb.2010.09.069}{265--271},
\href{http://arxiv.org/abs/1009.1643}{{\tt arXiv:1009.1643 [hep-th]}}

\bibitem{Brink:1976bc}
L.~Brink, J.~H. Schwarz  and J.~Scherk, \emph{{Supersymmetric Yang-Mills
  theories}}, Nucl.Phys. {\bf B121} (1977)
\href{http://dx.doi.org/10.1016/0550-3213(77)90328-5}{77}

\bibitem{Gliozzi:1976qd}
F.~Gliozzi, J.~Scherk  and D.~I. Olive, \emph{{Supersymmetry,
supergravity
  theories and the dual spinor model}}, Nucl.Phys. {\bf B122} (1977)
\href{http://dx.doi.org/10.1016/0550-3213(77)90206-1}{253--290}

\bibitem{deRoo:1984gd}
M.~de~Roo, \emph{{Matter coupling in $N=4$ supergravity}}, Nucl.Phys.
{\bf
  B255} (1985)
\href{http://dx.doi.org/10.1016/0550-3213(85)90151-8}{515}

\bibitem{Behrndt:2001ab}
K.~Behrndt, E.~Bergshoeff, D.~Roest  and P.~Sundell, \emph{{Massive
dualities
  in six dimensions}}, Class.Quant.Grav. {\bf 19} (2002)
  \href{http://dx.doi.org/10.1088/0264-9381/19/8/309}{2171--2200},
\href{http://arxiv.org/abs/hep-th/0112071}{{\tt arXiv:hep-th/0112071
[hep-th]}}

\bibitem{Carrasco:2013qia}
J.~J.~M. Carrasco and R.~Kallosh, \emph{{Hidden supersymmetry may imply
duality
  invariance}},
\href{http://arxiv.org/abs/1303.5663}{{\tt arXiv:1303.5663 [hep-th]}}

\bibitem{Kuzenko:2005wh}
S.~M. Kuzenko and S.~A. McCarthy, \emph{{On the component structure of
$N=1$
  supersymmetric nonlinear electrodynamics}}, JHEP {\bf 0505} (2005)
  \href{http://dx.doi.org/10.1088/1126-6708/2005/05/012}{012},
\href{http://arxiv.org/abs/hep-th/0501172}{{\tt arXiv:hep-th/0501172
[hep-th]}}

\bibitem{Liu:2010sk}
H.~Liu, H.~Luo, M.~Luo  and L.~Wang, \emph{{Leading order actions of
goldstino
  fields}}, Eur.Phys.J. {\bf C71} (2011)
  \href{http://dx.doi.org/10.1140/epjc/s10052-011-1793-0}{1793},
\href{http://arxiv.org/abs/1005.0231}{{\tt arXiv:1005.0231 [hep-th]}}

\bibitem{Zheltukhin:2010xr}
A.~Zheltukhin, \emph{{On equivalence of the Komargodski-Seiberg action
to the
  Volkov-Akulov action}},
\href{http://arxiv.org/abs/1009.2166}{{\tt arXiv:1009.2166 [hep-th]}}

\bibitem{Kuzenko:2010ef}
S.~M. Kuzenko and S.~J. Tyler, \emph{{Relating the Komargodski-Seiberg
and
  Akulov-Volkov actions: Exact nonlinear field redefinition}}, Phys.Lett. {\bf
  B698} (2011)
  \href{http://dx.doi.org/10.1016/j.physletb.2011.03.020}{319--322},
\href{http://arxiv.org/abs/1009.3298}{{\tt arXiv:1009.3298 [hep-th]}}

\bibitem{Freedman:2012zz}
D.~Z. Freedman and A.~Van~Proeyen, {\em Supergravity}.
\newblock Cambridge University Press,
2012.
\newblock

\end{thebibliography}

\providecommand{\href}[2]{#2}\begingroup\raggedright\endgroup

\end{document}